\documentclass[authoryear,preprint,12pt]{elsarticle}
\usepackage{graphicx}
\usepackage{eufrak}
\usepackage{natbib}
\usepackage{amsmath}
\journal{Journal of Computational Physics}
\newcommand{\parz}[1]{\frac{\partial #1}{\partial z}}
\newcommand{\fulz}[1]{\frac{d #1}{d z}}
\begin{document}
\begin{frontmatter}
\title{First Results From New 3D Spectral Simulations Of Anelastic Turbulent Convection}
\author[labKalo]{Kaloyan Penev}
\author[labJoe]{Joseph Barranco}
\author[labSass]{Dimitar Sasselov}
\address[labKalo]{60 Garden St., M.S. 10, Cambridge, MA 02138}
\address[labJoe]{1600 Holloway Avenue, San Francisco, CA 94132-4163}
\address[labSass]{60 Garden St., M.S. 16, Cambridge, MA 02138}
\begin{abstract}
We have adapted the anelastic spectral code of 
\citet{Barranco_Marcus_06} to simulate a turbulent convective layer with the 
intention of studying the effectiveness of turbulent eddies in dissipating 
external shear (e.g. tides). We derive the anelastic equations, show the
time integration scheme we use to evolve these
equations and present the tests we ran to confirm that our code does what we
expect. Further we apply a perturbative approach to find an approximate
scaling of the effective eddy viscosity with frequency, and find that it is in
general agreement with an estimate obtained by applying the same procedure to
a realistic simulation of the upper layers of the solar convective zone.
\end{abstract}
\end{frontmatter}

\section{Introduction}
\label{sec: intro}
	Dissipation of stellar tides and oscillations is often considered to be
	mainly due to the turbulent flow in their convective zones. Usually the
	effects of the turbulent flow are parametrized by some sort of
	effective viscosity coefficient. Clearly the situation is not as simple
	as that and the usual ``fix'' is to allow this viscosity coefficient to
	depend on the perturbation being dissipated, most notably its frequency
	and perhaps direction of the shear it creates. 

	Completely analytical treatments start by assuming a Kolmogorov spectrum
	for the turbulent flow and combine it with some prescription for the
	effectiveness of eddies in dissipating perturbations of the given
	period. Since Kolmogorov turbulence is isotropic the direction of shear
	is unimportant in such prescriptions. Two such prescriptions have been
	used.

	The first, proposed by \citet{Zahn_66, Zahn_89}, states that when the 
	period of the
	perturbation ($T$) is shorter than the turnover times ($\tau$) of some
	eddies, their dissipation efficiency should be proportional to the
	fraction of a churn they manage to complete in half a perturbation period,
	in other words, the effective viscosity coefficient scales like:
	\begin{equation}
		\nu = \nu_{max} \min\left[ \left(\frac{T}{2\tau}\right),
		1\right].
	\end{equation}

	The second prescription is due to \citet{Goldreich_Nicholson_77} and 
	\citet{Goldreich_Keely_77}. They argue that eddies with turnover times
	much bigger than the period of the perturbation will not contribute
	appreciably to the dissipation, and hence the effective viscosity should
	be dominated by the largest eddies with turnover times $\tau<T/2\pi$. 
	Then the Kolmogorov prescription of turbulence predicts that
	the effective viscosity will scale as:
	\begin{equation}
		\nu = \nu_{max} \min\left[\left( \frac{T}{2\pi\tau}\right)^2,
		1\right]
	\end{equation}

	Zahn's prescription has been tested against  tidal circularization 
	times for binaries containing a giant star
	\citep{Verbunt_Phinney_95}, and is in general agreement with
	observations. Also Zahn's prescription is in better agreement with
	observed tidal dissipation of binary stars in clusters 
	\citep{Meibom_Mathieu_05} and with the location of the red edge of the Cepheid 
	instability strip \citep{Gonczi_82}.

	The less efficient prescription has been used successfully by 
	\citet{Goldreich_Keely_77}, \citet{Goldreich_Kumar_88} and
	\citet{Goldreich_Kumar_Murray_94} to develop a theory for the damping of
	the solar $p$-modes. In this case the more effective dissipation would
	require dramatic changes in the excitation mechanism in order to
	explain the observed amplitudes. 

	Finally \citet{Goodman_Oh_97} developed a perturbative
	derivation of the convective viscosity, which for a Kolmogorov 
	scaling, gives a result that is
	closer to the less efficient Goldreich \& Nicholson viscosity than it 
	is to Zahn's. While providing a firmer theoretical
	basis for the former scaling, this does not resolve the
	observational problem of insufficient tidal dissipation.

	The development of 2D and 3D simulations of solar convection hint at a
	possible resolution of this problem. The convective flow that these
	simulations predict is fundamentally very different from the assumed
	Kolmogorov turbulence 
	\citep{Sofia_Chan_84, Stein_Nordlund_89, Malagoli_Cattaneo_Brummell_90}. 
	The first major difference is that
	the frequency power spectrum of the velocity field is much flatter than
	the Kolmogorov power spectrum, and hence one might expect that the 
	dissipation will 
	decrease significantly slower as frequency increases relative to the
	Kolmogorov case. Another major difference is that the velocity field is
	no longer isotropic and hence one would expect it to react differently
	to shear in different directions. That is, if we would use an
	effective viscosity coefficient, it should be a tensor and not a scalar
	quantity.

	As a first step in investigating that possibility 
	\citep{Penev_Sasselov_Robinson_Demarque_07}, 
	we applied the perturbative approach developed by \citet{Goodman_Oh_97} 
	to a numerical model of solar convection (Robinson et al. 2003) to find 
	the scaling of the components of the effective viscosity tensor with 
	frequency. Somewhat unexpectedly, we found that the scaling closely 
	follows Zahn's prescription, even though when the same approach is 
	applied to Kolmogorov turbulence it gives results closer to those of 
	Goldreich and collaborators.

	That left a lot of questions unanswered. For example, are 
	the effects of turbulence anything at all like that of molecular
	viscosity, and is taking only the lowest order term in a power
	series expansion of the energy dissipation rate an acceptable
	approximation? The perturbative approach also assumes that the 
	spatial scale of the external shear being dissipated is large compared
	to all convective scales, and hence can be assumed a linear function of 
	position, but it would be interesting to see how quickly dissipation 
	efficiency is lost as the spatial scale  of the external shear 
	decreases.

	To tackle these questions we adapted the anelastic spectral code 
	of \citet{Barranco_Marcus_06} to simulate a convectively 
	unstable box in which we are able to place external, time dependent 
	shear as part of  the evolution equations and observe the effects of 
	the turbulent convective flow directly. 
	
	In order to make the code
	suitable to simulate convection we added heat diffusion,
	that allows for the supply the heat that will drive the convective
	motions. This in turn necessitated that we allow for a more general
	background state slightly modified treatment of the pressure and the
	introduction of temperature boundary conditions. In addition we added an
	external forcing term and removed the Kepplerian shear and the
	accompanying it shearing coordinates.

	The resulting code is ideal for our purposes for the following reasons:
	\begin{enumerate}
		\item{The anelastic approximation means no shock or sound waves
		can exist and hence the time steps can be much larger than for 
		a fully compressible code, and yet the anelastic approximation 
		allows phenomena such as convection to occur unlike in 
		completely incompressible codes. This allows us to get much 
		closer to the actual time scales of interest (usually of order 
		hours or days) than a fully compressible code would. The
		Boussinesq approximation has those same properties, however it
		is not suitable for simulating stratified convection where the
		buoyancy does not depend entirely on the temperature.}

		\item{Spectral codes are more efficient at simulating turbulent
		flows than finite difference codes because their spatial
		accuracy is exponential rather than a power law, so they are
		capable of reliably reproducing the turbulent flow even at 
		modest resolution.}
	
		\item{The grid happens to have higher density near the top and 
		bottom of the box where it is needed in order to resolve the 
		boundary layers that develop there due to the boundary 
		conditions we impose.}
	\end{enumerate}

	In this paper we present the details of the modified code as well as a
	perturbative estimate of the efficiency of turbulent viscosity. In
	section \ref{sec: equations} we derive the anelastic approximation to 
	the Navier-Stokes equations modified to include a time dependent shear.
	In section
	\ref{sec: numerics} we present the numerical algorithm. In section 
	\ref{sec: tests} we show the tests we ran to
	confirm that our implementation actually evolves the desired equations,
	and verify that the numerical scheme is indeed second-order accurate in
	time. Finally, in section \ref{sec: results} we use a perturbative
	calculation to get an estimate of the effective turbulent viscosity in 
	our simulated box. 

\section{The evolution equations}
\label{sec: equations}
We define upfront the following quantities describing the fluid and the flow:
\begin{center}
\begin{tabular}{l@{ -- }ll@{ -- }l}
	$p$ & pressure &
	$T$ & temperature\\
	$\rho$ & density &
	$\mathbf{v}$ & velocity\\
	$\theta$ & potential temperature &
	$C_p$ & constant pressure specific heat\\
	$R$ & ideal gas constant &
	$g$ & acceleration of gravity\\
	$\kappa$ & heat diffusion coefficient &
	$\mathbf{f}$ & some external forcing (e.g. tidal force) 
\end{tabular}
\end{center}

As was said before we do not actually evolve the fully compressible fluid
equations but rather their anelastic approximation (c.f. \citet{Bannon_96}). 
Under the anelastic assumption each flow variable is split into a sum
of a time-independent background component (denoted by an over bar) and a time 
varying perturbation (denoted by tilde over its symbol) superimposed on top of
that. 

The background variables satisfy the fully compressible equations with all x, y
and t derivatives set to zero. The background velocity is assumed to be zero as
well. The boundary conditions that complete these equations are that we require 
the background temperature on the top and bottom walls to have some fixed values
$T_l$ and $T_h$ respectively, and the pressure at the top wall to have some
value, $p_{top}$. Alternatively, we could fix the temperature gradients at the
boundary, which would set the flux that the simulated convective layer must
transport. Neither of those two options is exactly what the actual physical 
problem requires. We could base our boundary conditions or reliable physics 
if we were planning to simulate the entire convective zone. However, this is 
impossible to do at the resolution required to study the turbulent dissipation, 
so we will ultimately be interested only in the flow that develops in the 
interior of the box and plan to exclude the regions near the top and bottom 
boundary from any analysis we do. For this reason the choice of the exact 
thermal boundary conditions will hopefully have little impact on our results.

For convenience, we define the following quantities:
\begin{eqnarray}
	\label{eq: K(z)}
	\mathcal{K}(z)&\equiv&\int_{-L_z/2}^z \frac{dz}{\kappa(z)}\\
	\label{eq: alpha}
	\alpha&\equiv&\frac{T_h-T_l}{\mathcal{K}(Lz/2)}
\end{eqnarray}
Where the $z$ coordinate has a value $-L_z/2$ at the bottom of the box and
$L_z/2$ at the top of the box.

The background variables that emerge as the solution to fully compressible 
equations are:
\begin{eqnarray}
	\bar{T}(z)&=&T_h-\alpha\mathcal{K}\\
	\bar{\rho}(z)&=&\frac{p_{top}}{R T_l} \exp\left[
	\int_z^{Lz/2}\frac{1}{\bar{T}}
	\left(\frac{g}{R}-\frac{\alpha}{\kappa}\right)dz\right]\\
	\bar{p}&=&\bar{\rho}R\bar{T}\\
	\bar{\theta}&=&\bar{T}\left(\frac{p_0}{\bar{p}}\right)^\frac{R}{C_p}
\end{eqnarray}
With those definitions the anelastic equations governing the evolution of the
perturbation variables become:
\begin{eqnarray}
	\label{eq: incompressibility anel}
	\nabla\cdot \bar{\rho} \mathbf{v}&=&0\\
	\label{eq: momentum anel}
	\frac{\partial \mathbf{v}}{\partial t} &=&  
	\mathbf{v}\times\mathbf{\omega}
	- \nabla \widetilde{h} +
	\frac{\widetilde{\theta}}{\bar{\theta}}g\hat{\mathbf{z}}+\mathbf{f}\\
	\label{eq: heat transfer anel}
	\frac{\partial \widetilde{\theta}}{\partial t} &=&
	-v_z\frac{d\bar{\theta}}{dz} -
	\mathbf{v}\cdot\nabla\widetilde{\theta}
	+ \frac{\bar{\theta}}{C_p \bar{T}\bar{\rho}}\nabla\cdot\left(\kappa\nabla \widetilde{T}\right)\\
	\label{eq: enthalpy anel}
	\widetilde{h}&\equiv&\frac{\widetilde{p}}{\bar{\rho}}+\frac{\mathbf{v}^2}{2}\\
	\label{eq: state anel}
	\frac{\widetilde{p}}{\bar{p}}&=&\frac{\widetilde{\rho}}{\bar{\rho}}
	+ \frac{\widetilde{T}}{\bar{T}}\\
	\label{eq: theta anel}
	\frac{\widetilde{\theta}}{\bar{\theta}} &=&
	\frac{\widetilde{p}}{\bar{\rho}gH_\rho} -
	\frac{\widetilde{\rho}}{\bar{\rho}}
\end{eqnarray}
Where we have introduced an enthalpy $\widetilde{h}$ (defined by equation 
\ref{eq: enthalpy anel}), a density scale height $H_\rho \equiv
-\left(\frac{d\ln\bar{\rho}}{dz}\right)^{-1}$ and the vorticity
$\mathbf{\omega}\equiv\mathbf{\nabla}\times\mathbf{v}$.
Equation \ref{eq: theta anel} is not the correct linearized equation for the
potential
temperature. The denominator of the pressure term should have been $\gamma
\bar{p}$ instead of $\bar{\rho}gH_\rho$. This replacement was introduced by
\citet{Bannon_96}. He showed that this substitution is required to ensure that 
the
anelastic equations conserve energy. To define the time evolution completely we
need to add boundary conditions to the above equations. The boundary conditions
on the four vertical walls of the domain are set by the fact that we use Fourier
series expansion for the horizontal spatial dependence of all quantities, and
hence all quantities are naturally periodic in those directions. In
addition to that, we want the temperature at the top and bottom boundary to be
whatever we specify, and hence its perturbation should be zero. Also we require
that the top and bottom walls are impermeable, but with no friction. That means
that  we set $v_z$ to zero at the top and bottom boundary, and do not require
anything for $v_x$ and $v_y$.

The anelastic equations above obey a set of energy conservation equations for
the following definitions of the kinetic and thermal energies:
\begin{eqnarray}
	E_K &\equiv& \int_V \bar{\rho}\frac{\mathbf{v}^2}{2} dV
	\label{eq: KE}\\
	E_T &\equiv& \int_V C_p\bar{\rho} \bar{T}
	\frac{\widetilde{\theta}}{\bar{\theta}} dV
	\label{eq: TE}
\end{eqnarray}
Using the anelastic evolution equations we can express the rates of change of
these two energies to be:
\begin{eqnarray}
	\frac{dE_K}{dt} &=& \mathcal{E}_1 \label{eq: E evol first}\\
	\frac{dE_T}{dt} &=& -\mathcal{E}_1 + \mathcal{E}_2
\end{eqnarray}
Where sinks/sources are defined as:
\begin{eqnarray}
	\mathcal{E}_1 &\equiv& g\int_V \frac{\bar{\rho}}{\bar{\theta}}
	\widetilde{v} \widetilde{\theta} dV \\
	\mathcal{E}_2 &\equiv& \int \left(\left.\kappa\frac{\partial
	\widetilde{T}}{\partial z}\right|_\frac{L_z}{2} - 
	\left.\kappa\frac{\partial
	\widetilde{T}}{\partial z}\right|_{-\frac{L_z}{2}}\right) dx dy
	\label{eq: E evol last}
\end{eqnarray}

\section{Numerical Time Evolution}
\label{sec: numerics}
\subsection{Spectral Method}
The spectral representation for a flow variable (q) used in the code is given by:
\begin{equation}
	q(x,y,z,t)\approx
		\sum_{k=-\frac{Nx}{2}+1}^\frac{Nx}{2}
		\sum_{l=-\frac{Ny}{2}+1}^\frac{Ny}{2}
		\sum_{m=0}^{N_z} \hat{q}_{klm}(t)
		e^{i 2\pi k x / L_x}
		e^{i 2\pi l y / L_y}
		T_m(z)
\end{equation}
Where the vertical basis functions: $T_m(z)\equiv\cos \left(m\cos^{-1}
\frac{z}{L_z}\right)$, are Chebyshev polynomials. The spectral method we use
does almost all calculations in the wavenumber/Chebyshev space, except for
taking products of variables, which are done by first transforming back to
physical ($x,y,z$) space on a grid of collocation points, taking the product 
there and transforming back. For a more complete discussion of the spectral 
method used see \citet{Barranco_Marcus_06} (section 3 and 3.2). 

\subsection{Time integration}
\subsubsection{Advection Step}
The advection step is fully explicit. It uses variables from this and the 
previous time step to achieve second order accuracy. It first calculates the 
quantities:
\begin{eqnarray}
	\mathfrak{M} &\equiv& \mathbf{v}\times\mathbf{\omega} + 
	\frac{\widetilde{\theta}}{\bar{\theta}}g \hat{\mathbf{z}}\\
	\mathfrak{N} &\equiv& -v_z\frac{d\bar{\theta}}{dz} -
	\mathbf{v}\cdot\nabla\widetilde{\theta}
\end{eqnarray}

Then velocity and temperature are updated using:
\begin{eqnarray}
	\mathbf{v}^{N+\frac{1}{3}} &=& \mathbf{v}^N + \frac{\Delta t}{2}\left(
	3\mathfrak{M}^N - \mathfrak{M}^{N-1}\right)\\
	\widetilde{\theta}^{N+\frac{1}{3}}  &=& \widetilde{\theta}^N + \frac{\Delta t}{2}\left(
	3\mathfrak{N}^N - \mathfrak{N}^{N-1}\right)
\end{eqnarray}

\subsubsection{Hyperviscosity Step}
The purpose of this step is to suppress the highest modes both in the 
horizontal and vertical directions to 
avoid buildup of power there due to the truncation of the spectrum at some 
finite number of spectral coefficients. For finite difference codes a step of
this sort is unnecessary because there is some finite ``grid viscosity''
associated with the numerical scheme. In spectral codes there is no equivalent
effect that prevents the build up of power at the highest simulated wavenumber 
modes. We implement the hypervisocisty step exactly in the way described in
\citet{Barranco_Marcus_06}, suppressing each spectral coefficient by a factor
as follows:
\begin{eqnarray}
	\mathbf{v}^{N+\frac{2}{3}} &=& \mathbf{v}^{N+\frac{1}{3}} \exp\left[
		-\Delta t \left(\nu_\perp^{hyp} k_\perp^{2p} + \nu_z^{hyp}
		m^{2p}\right)\right] \\
	\theta^{N+\frac{2}{3}} &=& \theta^{N+\frac{1}{3}} \exp\left[
		-\Delta t \left(\nu_\perp^{hyp} k_\perp^{2p} + \nu_z^{hyp}
		m^{2p}\right)\right] 
\end{eqnarray}
Where $\nu_\perp^{hyp}$ and $\nu_z^{hyp}$ are some hyperviscosity 
coefficients, $p$ is an integer between 1 and 6.
$k_\perp^2\equiv k_x^2+k_y^2$ is the horizontal wavenumber and $n$ is the order
of the Chebyshev polynomial that this particular amplitude applies to.

\subsubsection{Pressure Step}
\label{sec: pressure step}
The pressure step basically ensures that the velocity at the end of the time 
step satisfies the anelastic constraint. In the numerical scheme we abandon the
enthalpy at a specific time as a variable, and instead use its average between
two consecutive time steps:
\begin{equation}
	\Pi^{N+1}\equiv\frac{1}{2}\left(\widetilde{h}^{N+1}+\widetilde{h}^N\right)
\end{equation}
We then achieve second order accurate time evolution by updating the velocity
according to:
\begin{equation}
	\mathbf{v}^{N+1} = \mathbf{v}^{N+\frac{2}{3}} - \frac{\Delta t}{2}
	\left(\mathbf{\nabla} \widetilde{h}^{N+1} + \mathbf{\nabla} 
	\widetilde{h}^{N}\right) = 
	\mathbf{v}^{N+\frac{2}{3}} - \Delta t\mathbf{\nabla}\Pi^{N+1}
\end{equation}

So imposing the anelastic constraint we get:
\begin{equation}
	0 = \mathbf{\nabla}\cdot\mathbf{v}^{N+1} + v_z^{N+1}\frac{d\log\bar{\rho}}{dz} = 
	\mathbf{\nabla}\cdot\mathbf{v}^{N+\frac{2}{3}} - \Delta t \nabla^2 \Pi^{N+1}
	+ v_z^{N+\frac{2}{3}}\frac{d\log\bar{\rho}}{dz} - 
	\Delta t\frac{\partial \Pi^{N+1}}{\partial z}\frac{d\log\bar{\rho}}{dz}
\end{equation}
Regrouping and imposing $\left.v_z\right|_{\pm\frac{Lz}{2}}=0$ we get the
differential equation for updating $\Pi$ with its boundary conditions:
\begin{eqnarray}
	\left[\nabla^2 + \frac{d\log\bar{\rho}}{dz}\frac{\partial}{\partial
	z}\right] \Pi^{N+1} &=&
	\frac{1}{\Delta t}\left(\mathbf{\nabla}\cdot\mathbf{v}^{N+\frac{2}{3}}+ 
	v_z^{N+\frac{2}{3}}\frac{d\log\bar{\rho}}{dz}\right)\\
	\left.\frac{\partial \Pi^{N+1}}{\partial z}\right|_{\pm L_z/2} &=& 
	\frac{1}{\Delta t}\left.v_z^{N+\frac{2}{3}}\right|_{\pm L_z/2}
\end{eqnarray}
For details of the implementation of this equation in our code see appendix
\ref{app: pressure step}

\subsubsection{Heat Diffusion Step}
The heat diffusion step updates the potential temperature according to:
\begin{equation}
	\widetilde{\theta}^{N+1}=\widetilde{\theta}^{N+\frac{2}{3}} + 
	\frac{\Delta t}{2} \frac{\bar{\theta} R \kappa}{C_p \bar{p}}
	\left[
	\nabla^2 + \fulz{\ln \kappa} \parz{}
	\right]\left(\widetilde{T}^N+\widetilde{T}^{N+1}\right)
\end{equation}
When we express $\widetilde{T}$ in terms of $\mathbf{v}$,$\widetilde{h}$ and
$\widetilde{\theta}$, this gives a second order differential equation for 
$\widetilde{\theta}^{N+1}$. In terms of the following definitions:
\begin{eqnarray}
	\mathfrak{P}&\equiv&-\frac{\alpha \bar{\theta}}{g\kappa \bar{T}} 
	\left[
	\nabla^2-\fulz{\ln\kappa}\parz{}+\frac{\kappa'^2}{\kappa^2}-\frac{\kappa''}{\kappa}
	\right]
	\left(\widetilde{h}-\frac{v^2}{2}\right)\\
	C_1&\equiv&\fulz{\ln\kappa} - \frac{2g}{C_p\bar{T}}\\
	C_2&\equiv&\frac{g}{C_p\bar{T}}\left(
		\fulz{\ln\bar{\theta}} - \fulz{\ln\kappa}\right)\\
	C_3&\equiv&-\frac{2C_p\bar{\rho}}{\kappa\Delta t}\\
	C_4&\equiv&\frac{\alpha\bar{\theta}}{g\kappa\bar{T}}
\end{eqnarray}
The equation we solve during this step is:
\begin{eqnarray}
	\left(\nabla^2 + C_1\parz{} + C_2 + C_3\right)
	\widetilde{\theta}^{N+1} &=& \left(\mathfrak{P}^{N+1} + \mathfrak{P}^N\right)
	\nonumber\\
	&&{}- \left(\nabla^2 + C_1 \parz{} + C_2\right)\widetilde{\theta}^N
	+ C_3\widetilde{\theta}^{N+\frac{2}{3}} \quad\quad\quad{}
	\label{eq: heat diff}
\end{eqnarray}
This is a second order equation for $\widetilde{\theta}^{N+1}$ so we need two
boundary conditions to make the solution unique. These come from the requirement
that the temperature perturbation on the boundary vanishes:
$\left.\widetilde{T}\right|_{\pm \frac{L_z}{2}} = 0$. This requires:
\begin{equation}
	\left.\widetilde{\theta}^{N+1}\right|_{\pm \frac{L_z}{2}} =
	C_4 \left(\widetilde{h}^{N+1} - \frac{\mathbf{v}^2}{2}\right)
	\label{eq: heat diff bc}
\end{equation}
So we see that the differential equation \ref{eq: heat diff} uses only the
value of $\Pi^{N+1}$, but the boundary conditions 
need the value of the enthalpy at the updated time. There are two options of how
to obtain this value. One is to keep track of the enthalpy from the very 
beginning and after each pressure step to update it as:
\begin{equation}
	\widetilde{h}^{N+1}=2\Pi^{N+1}-\widetilde{h}^{N}
\end{equation}
However, if the initial value of $\hat{h}$ is not perfectly set this
prescription will lead to oscillations in the value of the enthalpy at the top
and bottom boundaries. To illustrate this assume that initially we set
$\widetilde{h}^{0}$ to a value that is slightly higher than what it should be.
Since $\Pi^1$ is calculated without reference to the 
initial conditions it will have the correct value. 
This will lead to $\widetilde{h}^{1}$ being slightly smaller and so on. These
oscillations are then translated to the interior of the box
through their effect on the potential temperature variable. Further, we found
that these perturbations grow with time for all the cases we ran.

As a result instead of introducing the additional variable $\hat{h}$ with the
only purpose of getting temperature boundary conditions we use an approximation
to its value through $\Pi$:
\begin{equation}
	\widetilde{h}^{N+1}\approx\frac{1}{2}\left( 3\Pi^{N+1}-\Pi^N \right)
\end{equation}
These boundary conditions are then imposed by ignoring the differential equation
at the top and bottom wall (where it does not make sense any way) and computing
the values of $\widetilde{\theta}$ at those boundaries to satisfy the boundary 
conditions. For the implementation details of this step see appendix 
\ref{app: heat diffusion}

\section{Tests}
\label{sec: tests}
Since all the Fourier transforms and differential operators are computed in
exactly the same way as in \citet{Barranco_Marcus_06}, see section 4.1 of that
paper for a discussion of the performance of the code. 

Below we present various tests we ran to confirm that the code is indeed
evolving the equations described above, and that the numerical scheme employed
is indeed second order accurate in time. We repeat essentially all the tests 
that Barranco and Marcus ran to confirm their numerical scheme. 
\subsection{The background state}
\label{sec: background}
\begin{figure}[tb]
\begin{center}
	\Large{a)}\includegraphics[angle=-90, width=0.45\textwidth]{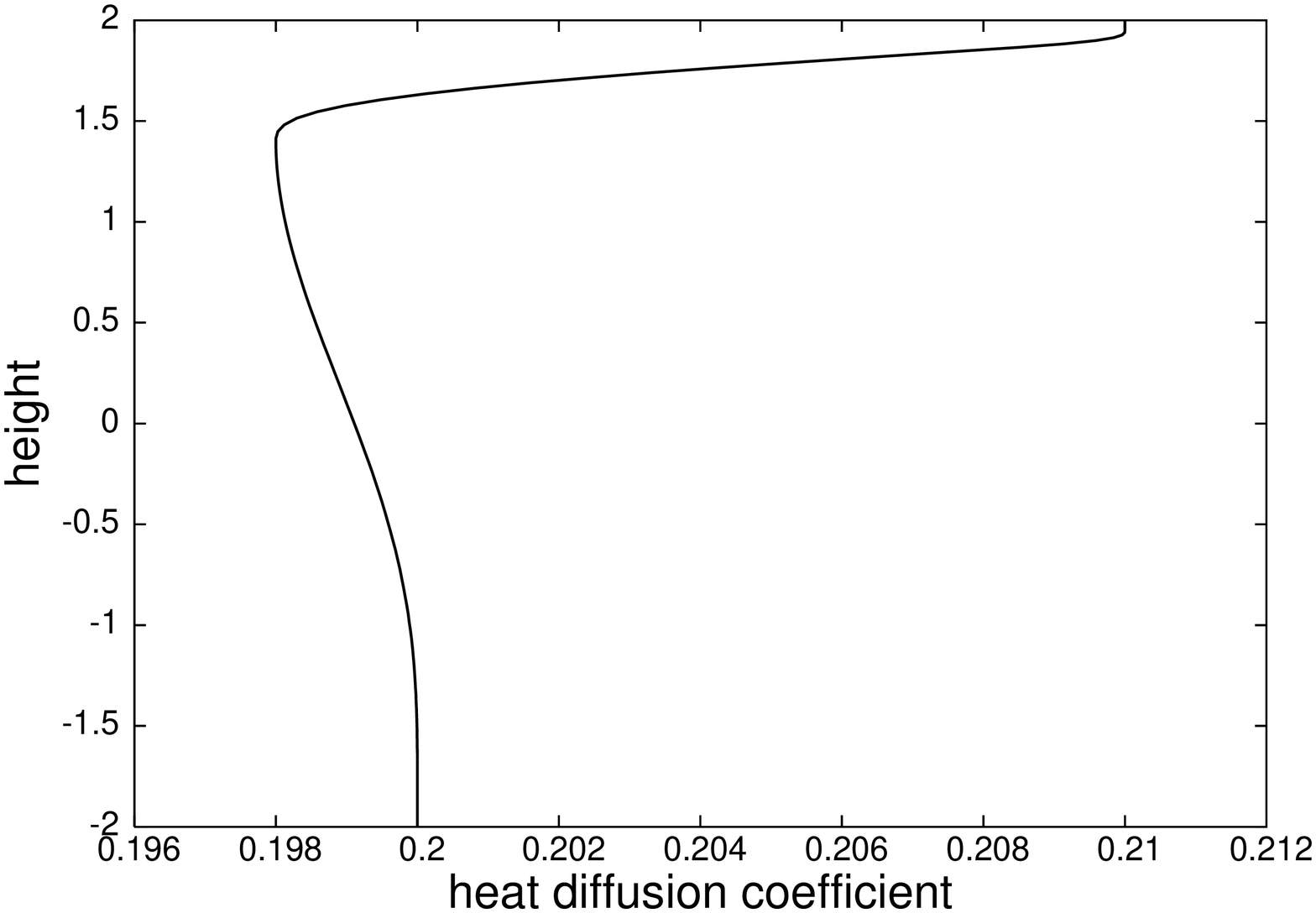}
	\Large{b)}\includegraphics[angle=-90, width=0.45\textwidth]{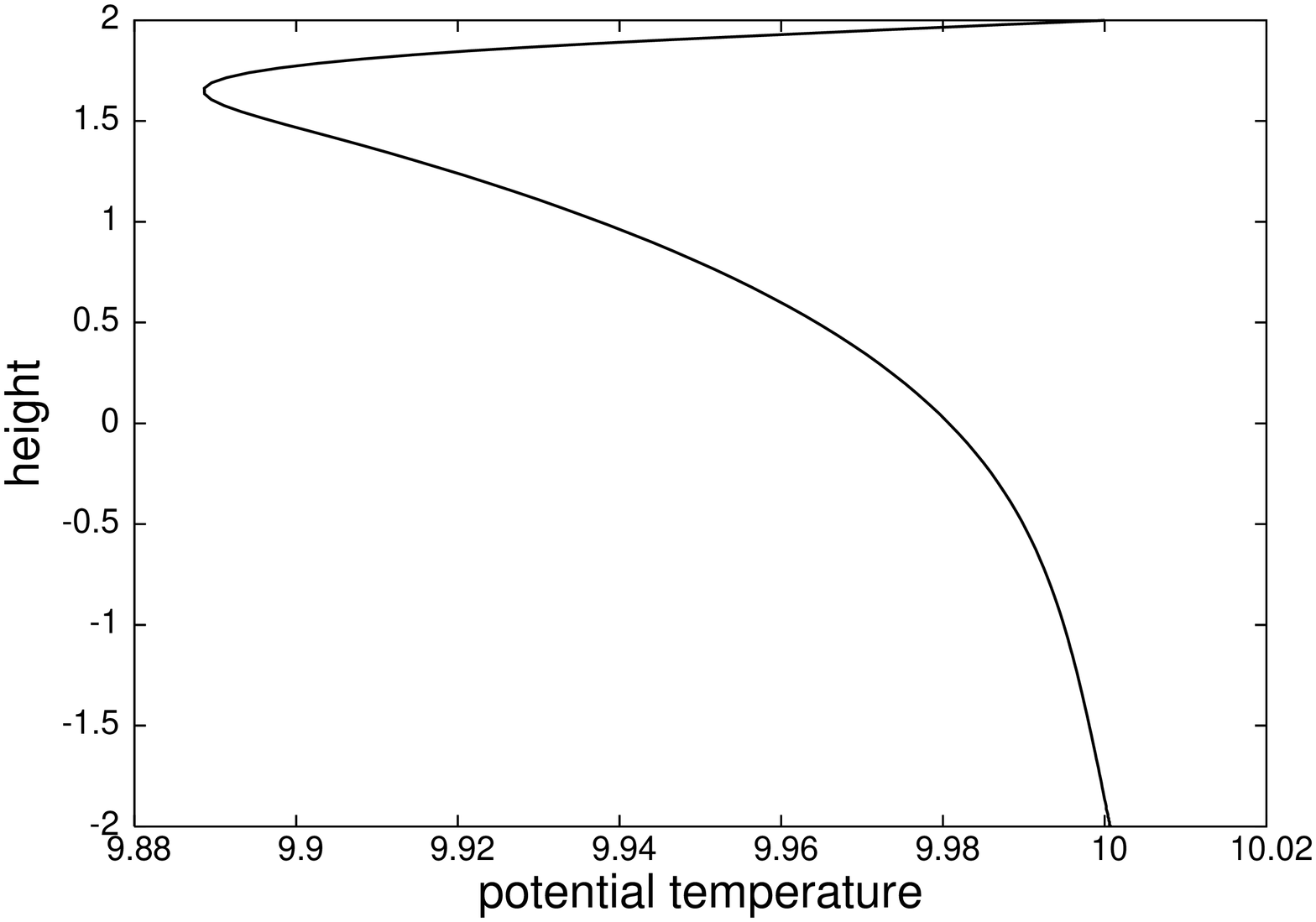}
	\Large{c)}\includegraphics[angle=-90, width=0.45\textwidth]{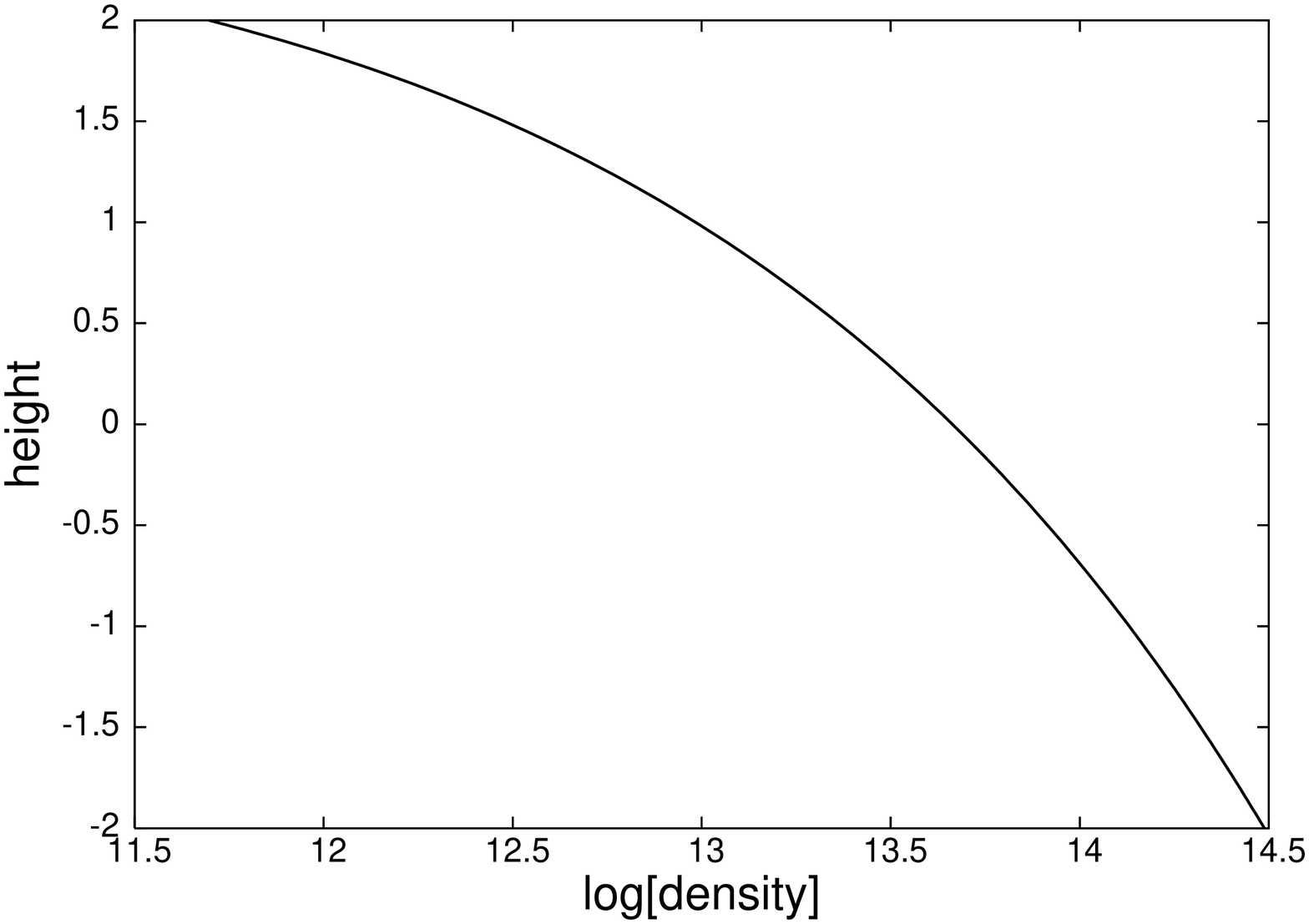}
	\Large{d)}\includegraphics[angle=-90, width=0.45\textwidth]{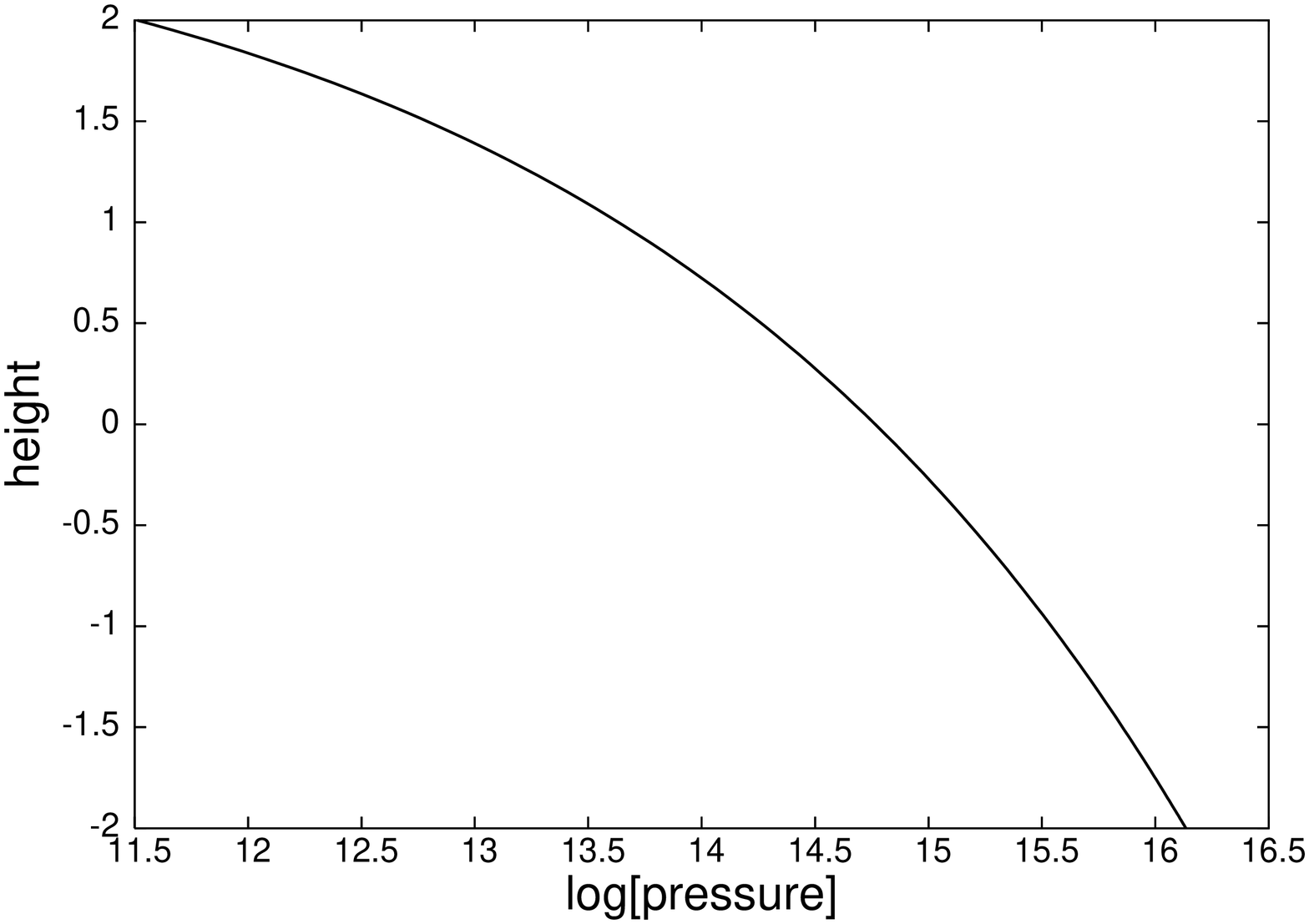}
	\caption{a) The heat diffusion coefficient ($\kappa$), 
	b) the background potential temperature ($\bar{\theta}$), 
	c) the background density ($\bar{\rho}$) and 
	d) the background pressure ($\bar{p}$)
	profiles.}
	\label{fig: kappa, theta, rho, p}
\end{center}
\end{figure}

Most of the tests we performed have the same background state, which is also the
state we used for the final convective box simulation. The only time we modified
the background was in order to simulate a convectively stable box and observe 
g-modes. Here we describe that background state. We use a self consistent 
non-dimensional set of units. Hence, the results apply for any choice of units.

The dimensions of the box are 4x4x4, with a resolution of 128 collocation points
per direction. 

Next we need to choose the 
vertical profile  of the heat diffusion coefficient. On one hand,
because we require the vertical velocity to vanish at the top and 
bottom boundaries, we need to set up a background that is stable to convection 
near those boundaries. On the other hand, we would like to simulate a 
top--driven convection, which is typical for stars with surface convection 
zones. To achieve that we need the most unstable stratification to occur near 
the top of the box. Further, since the heat diffusion step uses the heat
diffusion coefficient, and its first and second derivatives, we need our
expression for the heat diffusion coefficient to have a continuous second
derivative. 

To achieve all these requirements we construct the particular 
$\kappa(z)$ profile used in this paper from 6 separate pieces as follows:
\begin{equation}
	\kappa(z)=\left\{
	\begin{array}{l@{,}l}
		\kappa_0 & z< z_0 \\
		\kappa_1+\kappa_2\sin(k(z-z_0)) & z_0<z<z_1\\
		\kappa_3+\kappa_4 (z-z_1)\left[
			\left(\frac{z-z_1}{z_2-z_1}\right)^2-3
		\right] & z_1<z<z_2\\ 
		\kappa_5+\kappa_6(z-z_2)^2+\kappa_7(z-z2)^3+\kappa_8(z-z2)^4
								& z_2<z<z_3\\
		\kappa_9+\kappa_{10}*\sin\left(\pi\frac{z-z_3}{z_4-z_3}\right)
		+\kappa_{11}(z-z_3) & z_3<z<z_4\\
		\kappa_{12} & z_4<z
	\end{array}
	\right.
\end{equation}
Where the parameters $\kappa_i$, $i=1\ldots12$, $z_j$, $j=1\ldots4$ and k are
chosen to make $\kappa$, $\kappa'$ and $\kappa''$ continuous and to select the
shape of the curve. The shape of the curve used in this article was determined
from the following constraints:
\begin{itemize}
	\item The values that $\kappa(z)$ takes at the boundaries: $\kappa_0=20$
	and $\kappa_{12}=21$.
	\item The depth above which $\kappa(z)$ remains at its maximum value of
	$\kappa_{12}$: $z_4=1.95$.
	\item The depth at which $\kappa(z)$ has a minimum and the value at that
	minimum: $\kappa(z_2=1.4)=19.8$
	\item The depth below which $\kappa(z)$ is held constant at its bottom
	value of $\kappa_0$: $z_0=-1.8$
	\item The locations of the two inflection points: $z_1=0.2$ and
	$z_3=1.85$.
\end{itemize}
A plot of the resulting depth dependence of $\kappa(z)$ is 
presented in figure \ref{fig: kappa, theta, rho, p}a.

Next we need to choose values for the 
background temperature at the top ($T_{low}$) and bottom ($T_{high}$) of the 
box. Those are dictated by the requirement that the flow speeds should never 
approach the local speed of sound; otherwise, the anelastic approximation is no
longer an acceptable approximation and fully compressible equations should be 
used. 

Finally we need to choose values for 
the pressure at the top of the box ($p_{top}$), the external gravity ($g$), the
specific heat at constant pressure ($C_p$), and the value of the ideal gas
constant ($R$). These values, together with $T_{low}$ 
determine the pressure and density scale heights. Since we are interested in 
studying turbulence in a stratified medium we need to choose these values 
such that our box encompasses at least several pressure and density
scale heights. The particular values we chose were: 
\begin{eqnarray}
	p_{top}&=&1.0\times10^5\\
	g&=&2.74\\
	C_p&=&0.21\\
	R&=&8.317\times10^{-2}\\
	T_{low}&=&10.0\\
	T_{high}&=&62.37
\end{eqnarray}
We deliberately do not include any units in the above quantities, since
the simulated flow is independent of the choice of units.

The resulting background potential temperature,
pressure and density are presented in figures
\ref{fig: kappa, theta, rho, p}bcd. One
can see that the steepest negative slope of $\bar{\theta}$
occurs for heights between 1 and 1.5 units. Also we see we have significant
stratification. From figures \ref{fig: kappa, theta, rho, p}cd  
we see that the convective box encompasses
$2.8$ density scale heights and $4.6$ pressure scale heights.
\subsection{Energy}
We first verify that the energy like conserved quantities
defined in equation \ref{eq: KE} and equation \ref{eq: TE} evolve according to eqations
\ref{eq: E evol first} - \ref{eq: E evol last}. The initial conditions we used
for this test were that all components of the velocity were set to 0, and the
initial potential temperature contained random fluctuations in the lower $10\%$
of the spectral modes. We chose the time step to be much smaller than the
smallest absolute value of the buoyancy period, which in this case is an
imaginary quantity for most of the box. This way the time step is short both
compared to the growth rate of the instability near the middle of the box and
the g-mode period near the boundaries. \\

\begin{figure}
\begin{center}
	\includegraphics[width=0.49\textwidth]{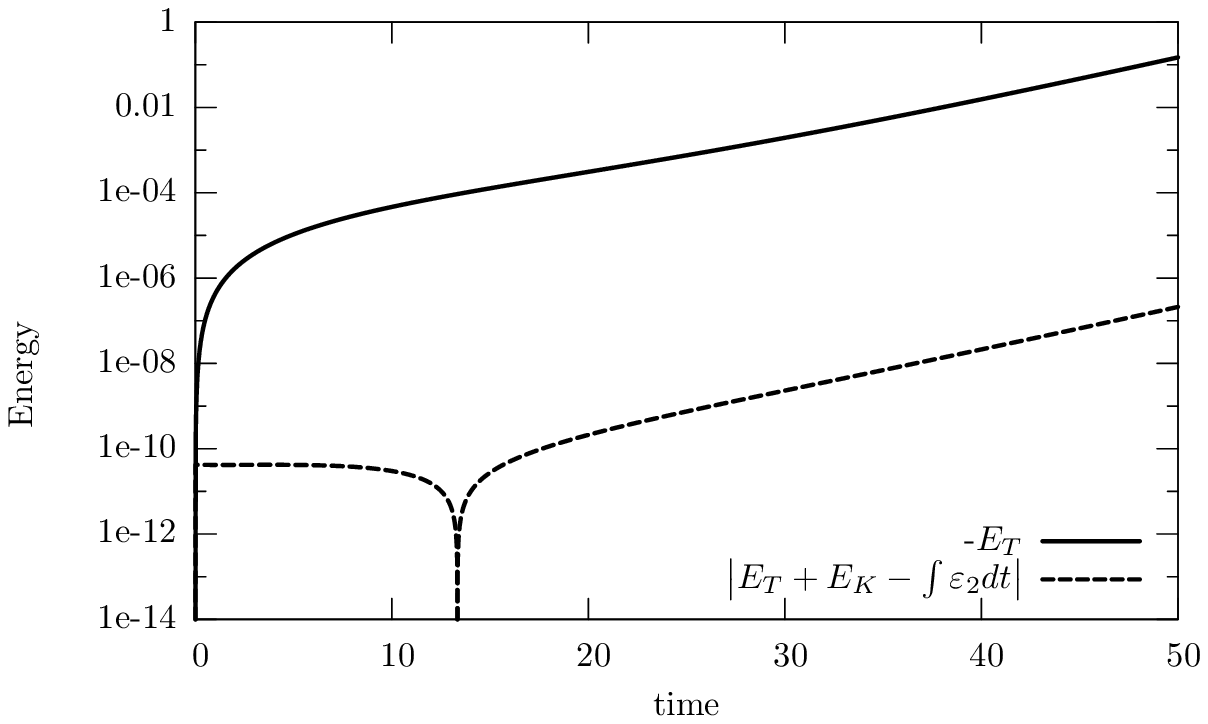}
	\includegraphics[width=0.49\textwidth]{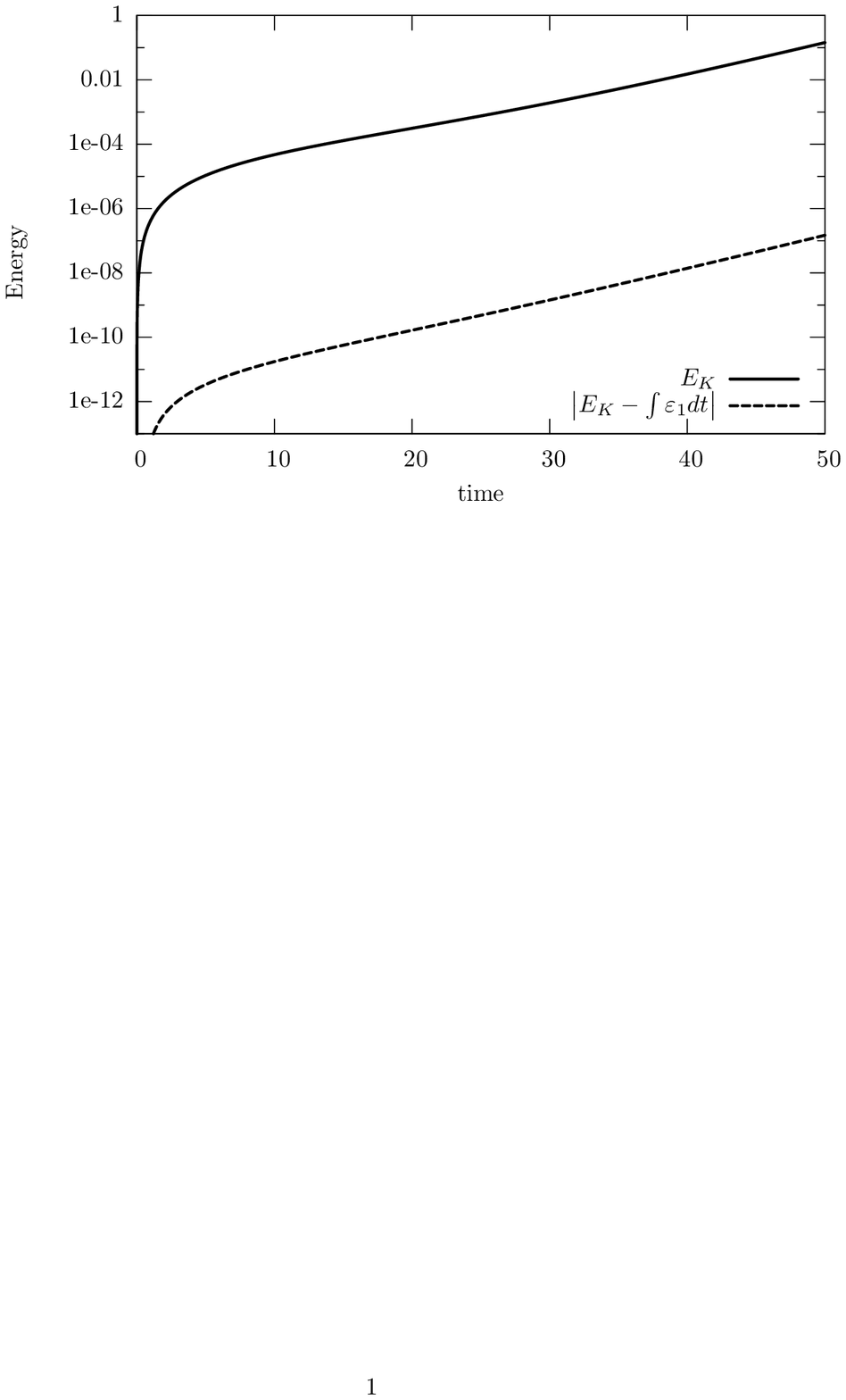}
	\caption{Energy conservation. Left: negative thermal energy -- 
	solid line, the absolute difference between the total energy calculated
	directly and as the integral of $\mathcal{E}_2$ -- dashed line; 
	Right : Kinetic energy (solid line), the absolute difference between the
	Kinetic energy calculated directly and as the integral of
	$\mathcal{E}_1$ -- dashed line}
	\label{fig: energy}
\end{center}
\end{figure}

We perform the test by running the code with no 
hyperviscosity. We output the kinetic and thermal energies as well as the rates
$\mathcal{E}_1$ and $\mathcal{E}_2$ at every time step. We then compute the time
integrals of $\mathcal{E}_1$ and $\mathcal{E}_2$ using Simpsons's method. This
is sufficient since the numerical evolution is only second order accurate in
time. In figure \ref{fig: energy} we show that equations 
\ref{eq: E evol first} -- \ref{eq: E evol last} are indeed satisfied to one 
part in a million. 

\subsection{Normal modes}
Another test we ran was checking that the normal modes of the linearized
equations of motion evolve as expected. We look for normal modes of the form:
\begin{equation}
	\widetilde{q}(x,y,z,t)= \hat{q}(k_x, k_y, z) e^{-i \omega t + i k_x x + i k_y y}
\end{equation}
For convenience define 
\begin{eqnarray}
	\hat{\tau}&\equiv& g\frac{\hat{\theta}}{\bar{\theta}} 
		\label{eq: tau_hat def}\\
	\omega_B^2&\equiv&g\frac{d\log\bar{\theta}}{dz}=
	\frac{g}{\bar{T}}\left(\frac{g}{C_p}-\frac{T_{high}-T_{low}}{L_z}\right)
		\label{eq: omega_B def}\\
\end{eqnarray}
In terms of equation \ref{eq: tau_hat def} and equation \ref{eq: omega_B def},
and after dropping all nonlinear terms, the anelastic equations (\ref{eq:
incompressibility anel} -- \ref{eq: theta anel}) simplify to
include only two variables $\hat{v}_z$ and $\hat{\tau}$:
\begin{eqnarray}
	-i\omega\left(I - \frac{D D_A}{k_\perp^2}\right) \hat{v}_z &=& 
	\hat{\tau} \\
	-i\omega\left[\hat{\tau} + 
	\frac{\kappa R \beta}{C_p \bar{p}}\left(D_A -D\right)\hat{v}_z\right] &=& 
	-\omega_B^2\hat{v}_z + \frac{\kappa}{C_p\bar{\rho}}\left[
	-k_\perp^2 + \frac{\beta}{\bar{T}} D + D^2 \right]\hat{\tau}
\end{eqnarray}
Where the operators $D$ and $D_A$ are defined in appendix A equations \ref{eq: D def}
and \ref{eq: D_A def}.

\begin{figure}
\begin{center}
	\includegraphics[width=1.0\textwidth]{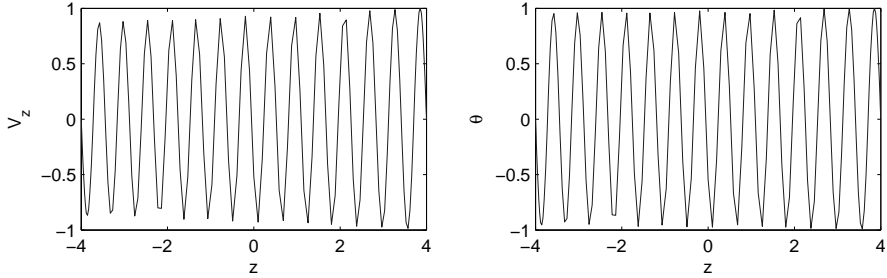}
	\caption{The shape of the normal modes we initilized the box with. On 
	the left is the $\mathbf{\hat{z}}$ component of velocity and on the 
	right is the potential temperature.}
	\label{fig: eigenmodes}
\end{center}
\end{figure}

In terms of $\hat{v}_z$ and $\hat{\tau}$ the remaining flow variables can be 
expressed as:
\begin{eqnarray}
	k_\perp^2 \hat{h} &=&i\omega D_A \hat{v}_z\\
	i\omega \hat{v}_x &=& i k_x \hat{h} \\
	i\omega \hat{v}_y &=& i k_y \hat{h} 
\end{eqnarray}
We calculated numerical solutions to the eigenmode equations (see figure 
\ref{fig: eigenmodes}) and supplied them as initial conditions with very low
amplitude to 
the code. We expect that for later times the evolution is done simply by
multiplying the initial amplitudes by $e^{-i\omega t}$. We ran the code with the
background already discussed above and a time step that was no larger than
$\frac{\pi}{250 \omega}$ for the mode in question. The comparison with the
simulated evolution of these eigenmodes is presented in figure \ref{fig:
eigenmode errors}. As we can see for large enough time steps the error scales as
the
square of the time step, which confirms that the code is indeed second order
accurate in time. For very small time steps the error deviates from that scaling
due to numerical round off. The minimal fractional error is much larger than the
numerical precision, because we have chosen the heat diffusion to be as small as
possible, which causes the matrix we invert during the heat diffusion step to have
values along the diagonal that are many orders of magnitude larger than the off
diagonal values, which causes the numerical roundoff to be amplified many times.
This also explains why the error in the potential temperature is largest (the
other varibles suffer from this only indirectly).

\begin{figure}[!tb]
\begin{center}
	\includegraphics[width=1.0\textwidth]{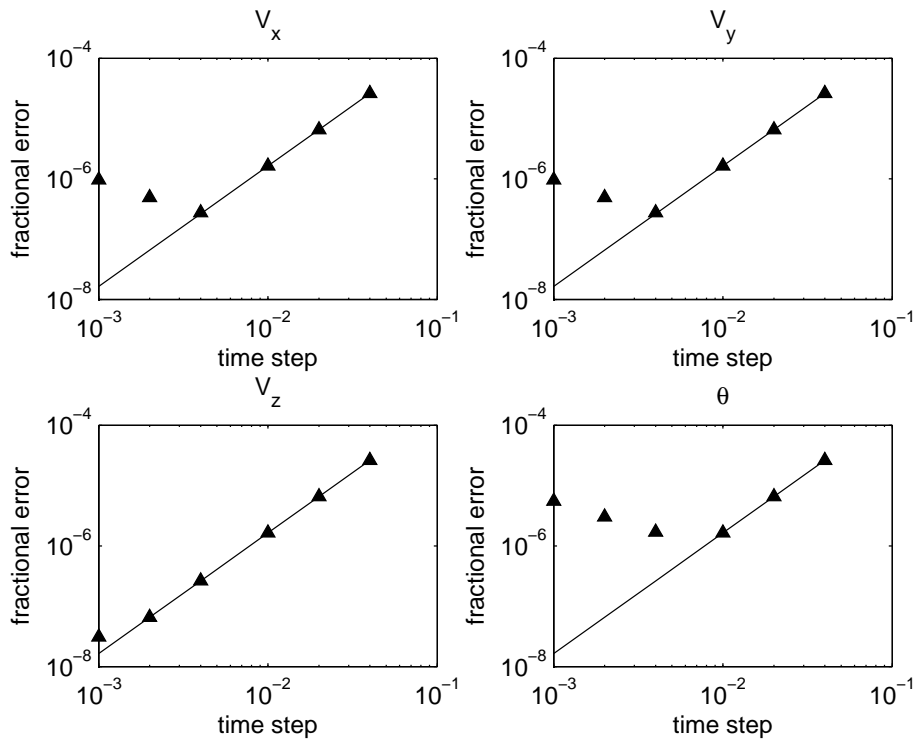}
	\caption{The maximum error in normal mode evolution, after a fixed
	 evolution time. The error is scaled by the
	maximum value of the quantity at time $t=0$. The solid line corresponds
	to quadratic scaling of the error with the time step.}
	\label{fig: eigenmode errors}
\end{center}
\end{figure}
\subsection{2D vortices}
\begin{figure}
\begin{center}
	\includegraphics[width=0.5\textwidth]{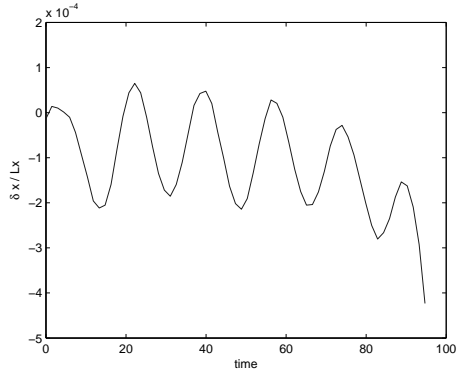}
	\caption{The difference between expected vortex position and the 
	average position of vorticity. The period
	of the oscillations corresponds to the rotational period of the vortices
	and the drop in the end is due to the fact that the vortices  are
	exiting the box on the right and hence a bit of them is appearing in the
	left, causing the average position of vorticity to move toward zero.}
	\label{fig: vortices}
\end{center}
\end{figure}

Another test we ran was initializing the box with a pair of circular columns of
vorticity running from the top to the bottom boundary. In physical space for
each column the vorticity was constant and in the $\hat{z}$ direction. 
The two columns had
opposite signs of vorticity so as to ensure that the total vorticity in the box
is zero, as required by the periodic boundary conditions. One expects that the 
two vortices will move parallel at a constant rate determined by the distance
between them and the magnitude of their vorticity. Because of the periodic
boundary conditions, having two vortices in our box is equivalent to having an
infinite number of vortices, copies of the two in the box. There is no
analytical expression of the infinite series for the velocity with which each 
vortex should move, but if the two vortices are far away from the walls as
compared to the distance between them we expect that their motion is at least
approximately that of the situation of only 2 vortices. The time step we chose
in this case was $0.01 t_{cross}$ (the expected time it would take the vortices
to cross half the box). We ran this test and verified that the rate at which the
vortices moved corresponded to the approximate analytical rate, see figure 
\ref{fig: vortices}.

\subsubsection{Convectively stable box}
\begin{figure}
\begin{center}
	\includegraphics[width=0.6\textwidth]{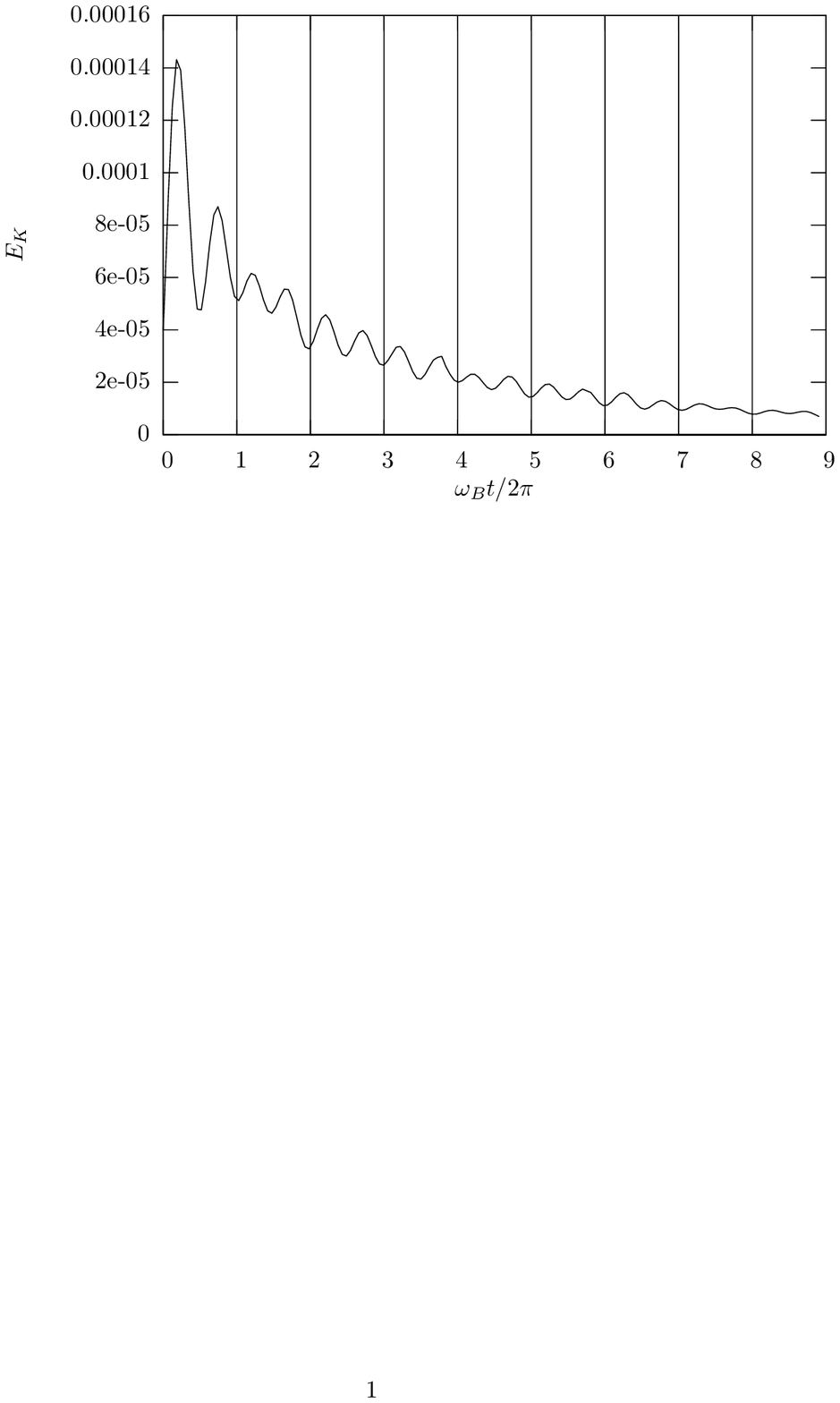}
	\caption{Kinetic energy as a function of time for a convectively stable
	box. The distance between every two consecutive vertical lines is the
	buoyancy period. The time is in units where the buoyancy period is 1.}
	\label{fig: convectively stable}
\end{center}
\end{figure}

The last test we ran was imposing a convectively stable stratification in the
box, and initializing with random temperature perturbations. In this case one
expects to see g-mode oscillations with a frequency given by the buoyancy
frequency $\omega_B^2\equiv g d\ln\bar{\theta}/dz$. So for this test we
needed the buoyancy frequency to be approximately constant throughout the box,
and much larger than the time step, but small enough to allow us to simulate
many buoyancy periods. So for this test we set the heat diffusion coefficient to
be constant throughout the box. Also we set the top and bottom temperatures to
be the same. This way the buoyancy frequency was independent of height. The
particular values for the parameters of this run were:
\begin{eqnarray}
	p_{top}&=&1.0\times10^5\\
	g&=&2.0\\
	C_p&=&0.2\\
	R&=&8.317e-2\\
	T_{low}&=&10\\
	T_{high}&=&10\\
	dt&=&0.002
\end{eqnarray}
This means that the buoyancy frequency is $\omega_B^2=2.0$, 
A plot of the kinetic energy
for this run is presented in figure \ref{fig: convectively stable}. It can be seen
that the kinetic energy goes through two cycles every period. This is because
the velocity has to go through one cycle, and the energy has a maximum every
time the velocity has a maximum or a minimum. The decay in the curve is caused
mostly by the heat diffusion smoothing out the perturbations over time. Since different
modes decay at different rates and they are coupled through the nonlinear terms
we have not shown an expected decay curve.
\section{Estimating the dissipation in an unstably stratified convective box}
\label{sec: results}
\subsection{Steady state flow}
\begin{figure}
\begin{center}
	\includegraphics[width=0.45\textwidth]{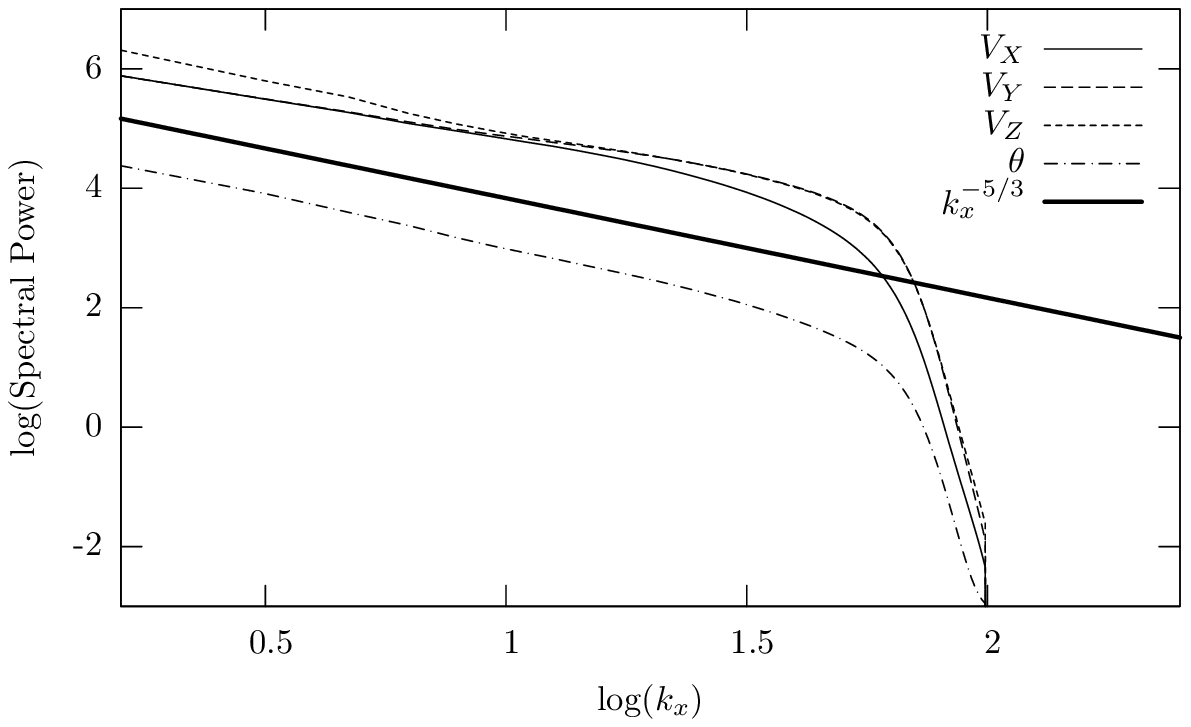}
	\includegraphics[width=0.45\textwidth]{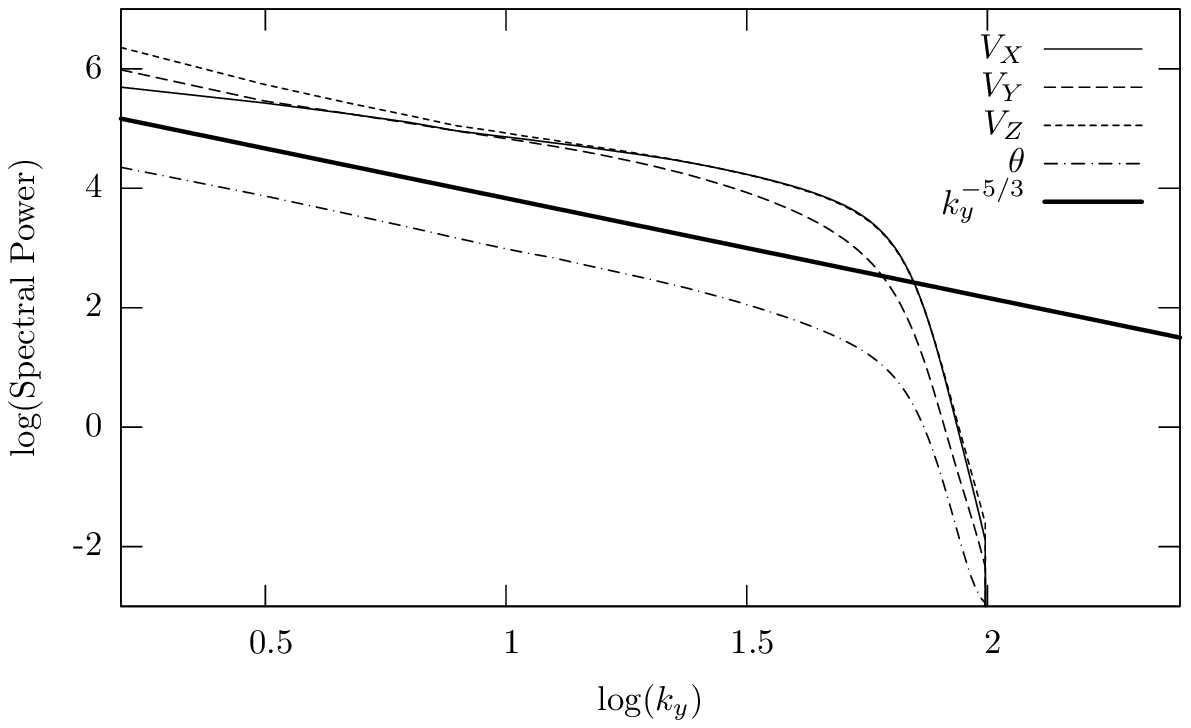}
	\includegraphics[width=0.45\textwidth]{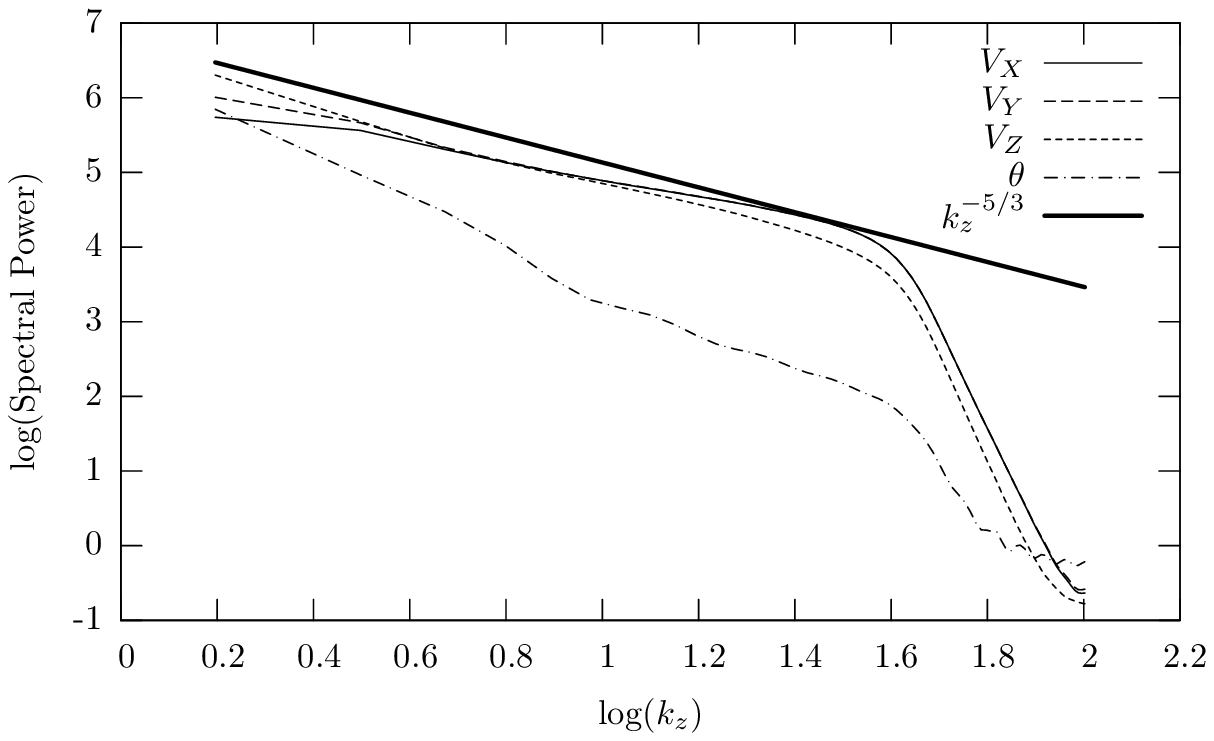}
	\includegraphics[width=0.45\textwidth]{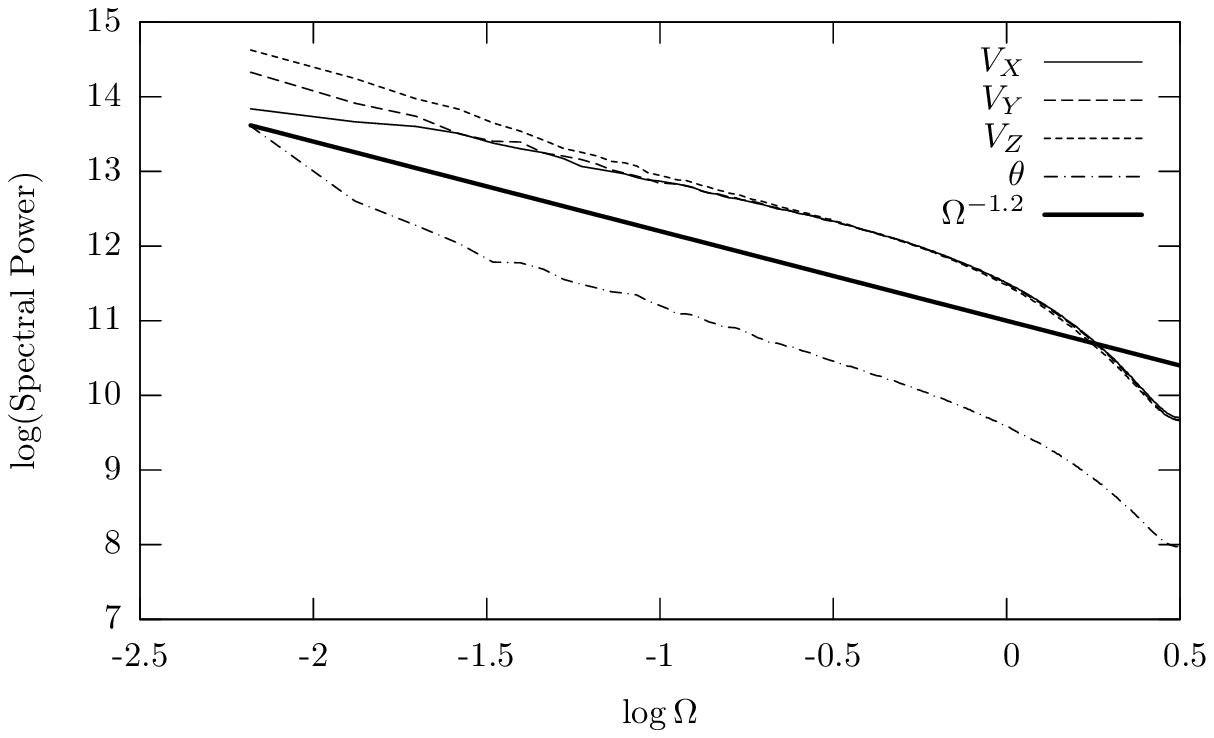}
	\caption{The (x: top left, y: top right, z: bottom left, time: bottom
	right) spectra of the 3 velocity components and the potential 
	temperature. The thick line in the spatial spectra plots corresponds to
	Kolmogorov scaling ($E_k\propto k^{-5/3}$). The thick line in the time
	spectra plot corresponds to the scaling we find for the effective
	viscosity in the next section.}
	\label{fig: fff spectra}
\end{center}
\end{figure}

Having confirmed that the code is solving the correct equations, we ran a box
with the background state described in section \ref{sec: background} for long 
enough to reach a steady--state flow. The criteria for having reached 
a steady--state flow were that the kinetic and thermal energies should stop drifting
systematically up or down (see figure \ref{fig: energy steady}). The oscillations
we see in the kinetic energy have a period close to the convective turnover time
of the box. We also want the
spatial spectra of the velocity and potential temperature to be steady to 
within a few percent. The vertical spectra can not be directly obtained 
from the output of the simulation since the collocation points of computational 
grid are not evenly spaced in the vertical direction. So we first had to 
re-sample to an evenly spaced grid and apply some window function in the
vertical direction before performing the discreet Fourier 
transform. Since in the vertical direction we simulate the Chebyshev series of 
the quantities the most natural way to re-sample to an even grid was to evaluate
this series for each of the new (evenly spaced) collocation points.
Fourier power spectra of the 3 velocity components and the potential temperature
are  presented in figure \ref{fig: fff spectra}, along with the scaling that
Kolmogorov statistics predict (spectral power $\propto k^{-5/3}$). The sharp
cutoff at high wavenumbers is related to the resolution of the box (using a box
of half the resolution produces a cutoff at half the wavenumber we see in figure
\ref{fig: fff spectra}).
\begin{figure}
\begin{center}
	\includegraphics[width=1.0\textwidth]{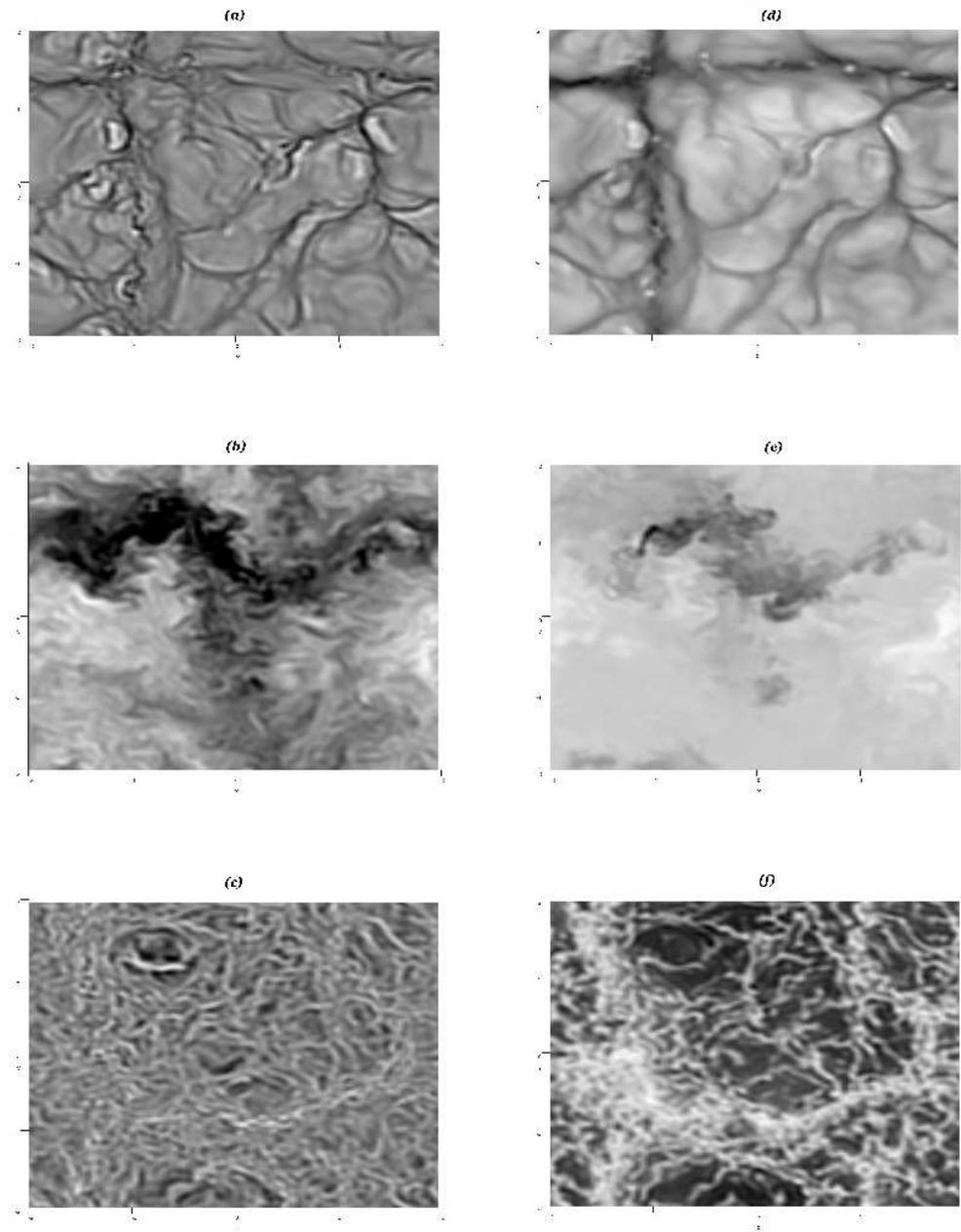}
	\caption{Horizontal cross sections of the vertical velocity component
	(left) and the total potential temperature (right) at three different
	heights of the box: $z=1.98$ (a,d), $z=0$ (b,e) and $z=-1.98$ (c,f)}
	\label{fig: hor sections}
\end{center}
\end{figure}
\begin{figure}
\begin{center}
	\includegraphics[width=1.0\textwidth]{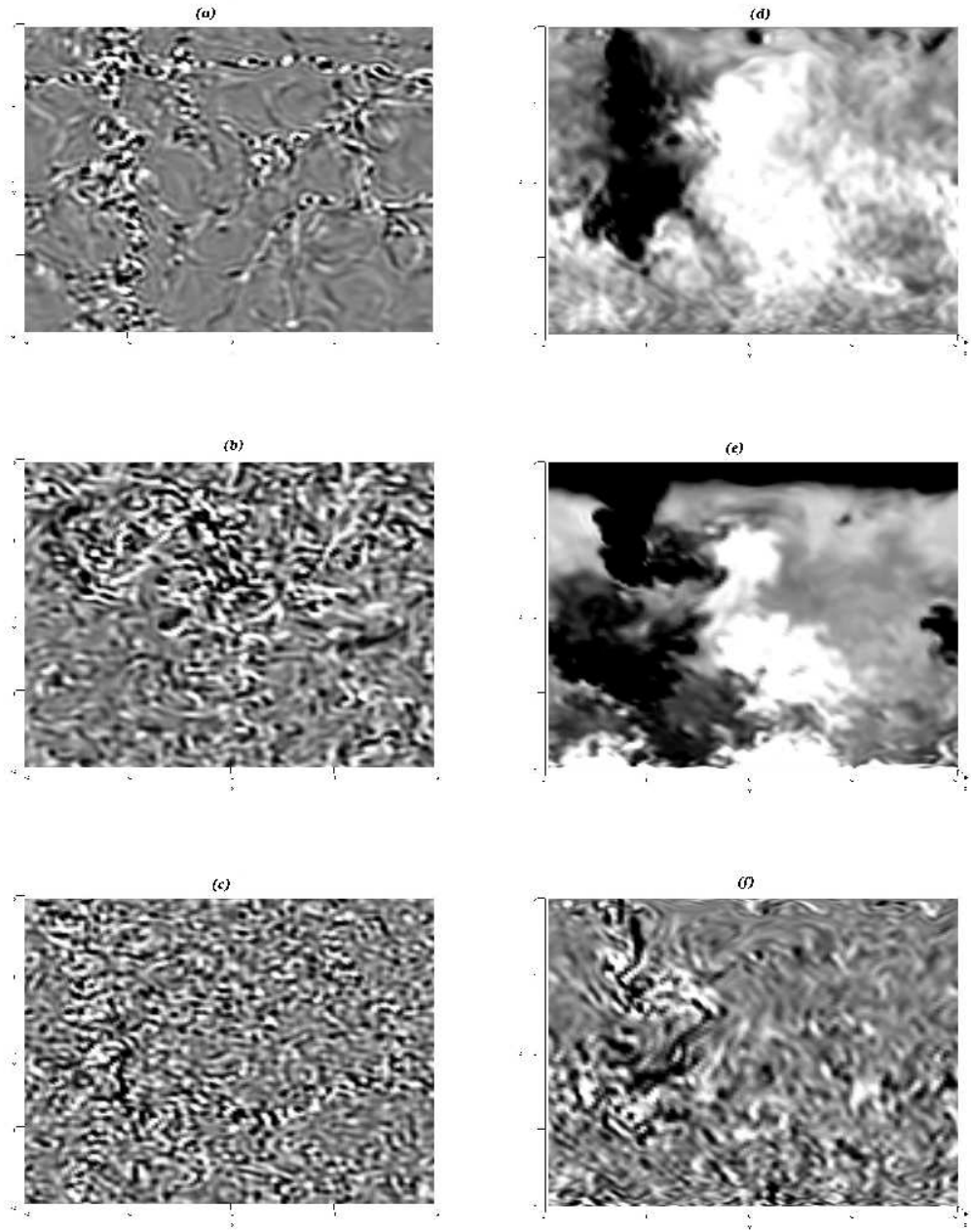}
	\caption{Horizontal cross sections of the vertical vorticity component
	(left panels) at heights:  $z=1.98$ (a), $z=0$ (b) and $z=-1.98$ (c). 
	Vertical cross sections (right panels) of the vertical velocity (d),
	the total potential temperature (e) and the vertical vorticity (f).}
	\label{fig: vort and vert sections}
\end{center}
\end{figure}
\begin{figure}[tb]
\begin{center}
	\includegraphics[width=0.6\textwidth]{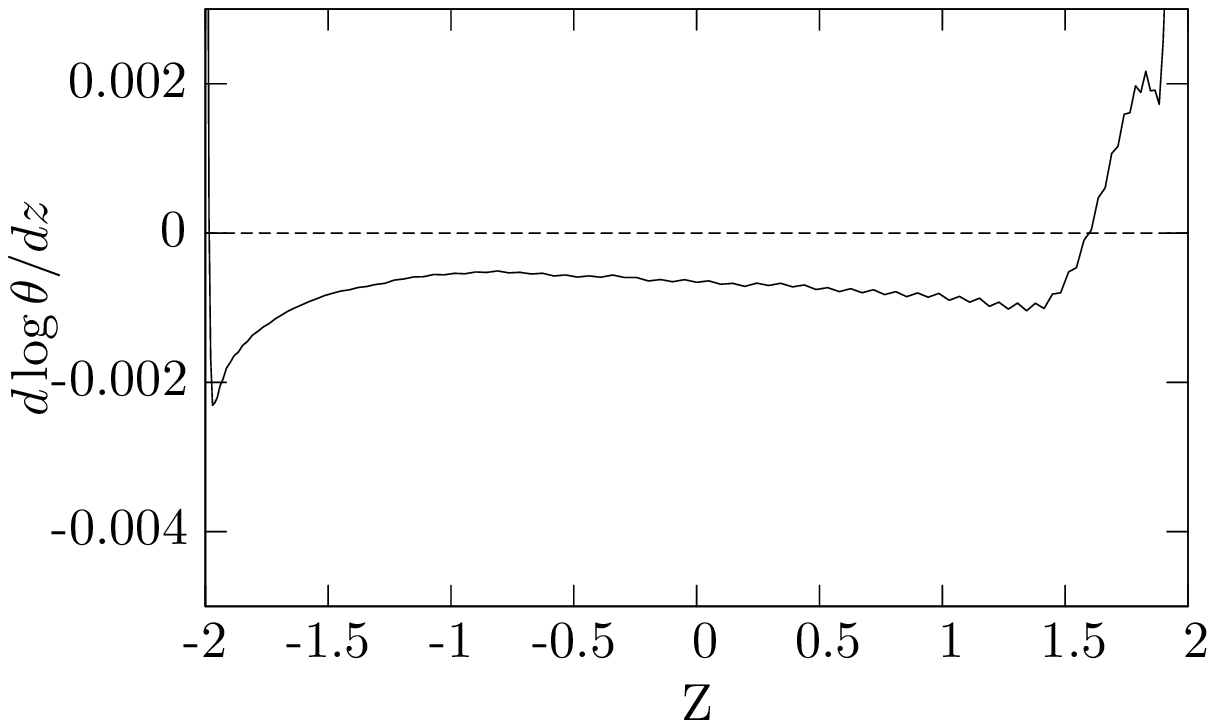}
	\caption{The average logarithmic gradient of the potential temperature.}
	\label{fig: dtheta_dz}
\end{center}
\end{figure}

Horizontal cross sections of $v_z$ and $\widetilde{\theta}+\bar{\theta}$ 
at 3 different heights of a typical steady state frame 
of the flow are presented in figure \ref{fig: hor sections}. The three heights we
chose were $z=1.98$, $z=0$ and $z=-1.98$ for a box in which the vertical 
coordinate runs from -2 to 2. Horizontal section of the $\hat{z}$ component of
the vorticity $\omega_z$ as well as typical vertical sections of the above
quantities are presented in figure \ref{fig: vort and vert sections}.

At the top of the box (figures \ref{fig: hor sections}a,d), near the top boundary, 
the flow exhibits a cellular pattern of narrow downflow lanes. Traces of the
cells are visible only in the top 5\% of the box, after that the flow organizes
itself into a pair of perpendicular downflow lanes that horizontally span the 
entire box.
With time, those lanes get distorted, break up and reform, but are well defined
for at least half the time, for most of the upper half of the box. 
They are generally parallel to the grid axis.
Similar patterns
were first observed by \citet{Porter_Woodward_00}. Their tests show that 
the lanes tend to align themselves with the periodic directions of the models,
and not with the grid axis per se (rotating the periodic directions at $45^o$
relative to the grid axis caused the lanes to rotate as well). So they conclude
this to be due to the small horizontal to vertical aspect ratio of the
simulations.

Around the middle of the box the pair of perpendicular lanes are still sometimes
visible, but they are a lot less persistent. Rather at those depths the flow
consists of one or two downflows, which take up less than a quarter of
the cross sectional area among a gentler upflow (figures \ref{fig: hor
sections}b,e).
For the lower half of the box, the asymmetry between up and downflows gradually
decreases as we get further down. Near the bottom boundary the large scale of
the flow disappears, until only small scale structure is left near the bottom
boundary (figures \ref{fig: hor sections}c,f).

Since we ran our simulations with the smallest possible value of the heat
diffusion coefficient we expect to have very efficient convection, hence we
would expect the z gradient of the potential temperature to be very close to
zero except near the impenetrable top and bottom boundaries. The averaged over
time and horizontal directions logarithmic gradient of
$\widetilde{\theta}+\bar{\theta}$ can
be seen in figure \ref{fig: dtheta_dz}. As we can see the scale height of $\theta$
is indeed more than two orders of magnitude larger than the box, as long as we
are not very close to the boundaries. The top of the box has a
local minimum of the entropy gradient and a significant convectively stable
region, so as to keep the large magnitude vertical flow from getting too close 
to the
boundary and causing numerical problems. We also see that another local minimum
of the entropy gradient has developed near the bottom of the box. This is
caused by the lower impenetrable wall. Not having a significantly thick
convectively stable region near the bottom of the box is acceptable, 
because the larger density of the fluid means that the flow velocities
are much smaller at the bottom than at the top of the box. 

\subsection{Lowest order perturbative expansion}
Ignoring the part of the run before steady state was reached we adapted the
\citet{Goodman_Oh_97} method as described in 
\citet{Penev_Sasselov_Robinson_Demarque_07, Penev_Sasselov_Robinson_Demarque_08}
to find a lowest order perturbative
expansion estimate of the energy transfer rate between a small amplitude
external forcing and the turbulent flow in our box, and respectively from that
we can derive the components of an effective viscosity tensor that 
reproduces the energy dissipation rate due to the turbulence. We assume 
forcing in the form of an external velocity field:
\begin{figure}[tb]
\begin{center}
	\includegraphics[angle=-90, width=0.49\textwidth]{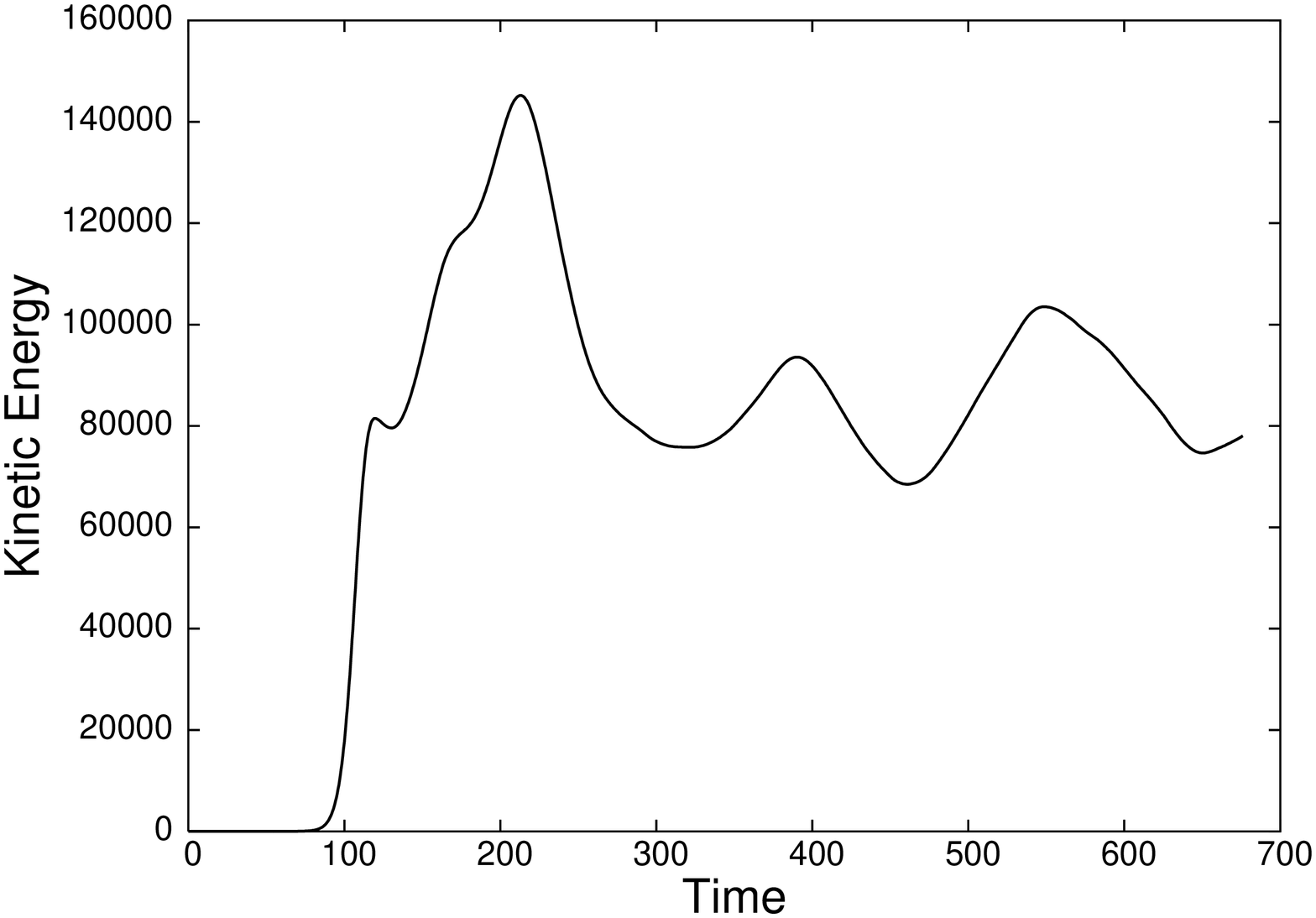}
	\includegraphics[angle=-90, width=0.49\textwidth]{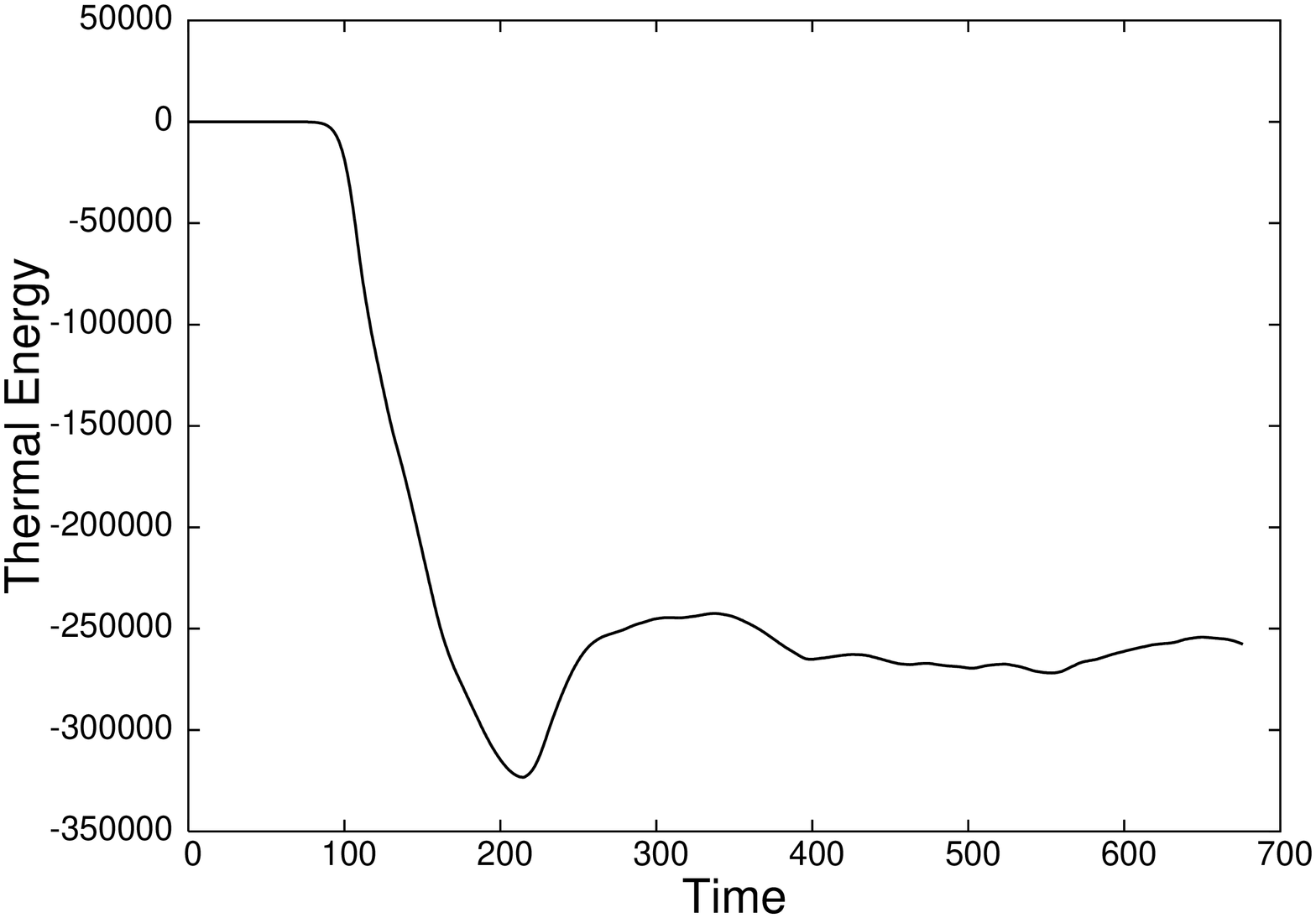}
	\caption{Kinetic (left) and thermal (right) energy content of the
	convective box used as one of the criteria for having reached a steady
	state. We decided steady state was reached at time time of 400.}
	\label{fig: energy steady}
\end{center}
\end{figure}
\begin{equation}
	\mathbf{V} = \mathbf{A}(t)\cdot \mathbf{x}
\end{equation}
\citet{Goodman_Oh_97} define two dimensionless parameters: the tidal strain
$\Omega^{-1} \left|\mathbf{A}\right|$, and $\left(\Omega
\tau_c\right)^{-1}$, where $\Omega$ is the frequency of the
perturbation and $\tau_c\equiv L_c/V_c$. The characteristic convective 
length scale is  $L_c$ and $V_c$  is the   
characteristic convective velocity. In the case of
hierarchical eddy structured convection $\tau_c$ is the largest eddy
turnover time.\\

We then take only the lowest nonzero term in the expansion of the secular rate
of change of the kinetic energy $\dot{\mathcal{E}}$ with respect to those two
parameters. \citep{Penev_Sasselov_Robinson_Demarque_08} adapt the
\citep{Goodman_Oh_97} formalism for discretely sampled data on a finite spatial
and temporal domain the resulting expression is:
\begin{eqnarray}
		\dot{\mathcal{E}}_{2,2}(\Omega=2\pi R/T)=\frac{2T}{\mathcal{N}^2 N_z}
		\sum_{\lambda,\mu,\nu,\nu'}
		Re\left\{
		\hat{\rho}^*_{\nu-\nu'}\left[
		\mathbf{\hat{v}}_{\lambda,\mu,\nu,-R}\cdot\mathbf{\widehat{A}}(\Omega) 
		\mathbf{P}_{\lambda,\mu,\nu'}
		\mathbf{\widehat{A}}(-\Omega)\cdot
		\mathbf{\hat{v}}^*_{\lambda,\mu,\nu',-R}
		\right]\right\}&&\nonumber\\
		-\frac{4}{\mathcal{N}^2 N_z} \sum_{\lambda,\mu,\nu,\nu'}
		\sum_{\rho\neq0}\frac{1}{\omega_{\rho}}Im\left\{
		\hat{\rho}^*_{\nu-\nu'}\left[
		\mathbf{\hat{v}}_{\lambda,\mu,\nu,\rho-R}\cdot\mathbf{\widehat{A}}(\Omega) 
		\mathbf{P}_{\lambda,\mu,\nu'}
		\mathbf{\widehat{A}}(-\Omega)\cdot\mathbf{\hat{v}}^*_{\lambda,\mu,\nu',\rho-R}
		\right]\right\}&&
	\label{eq: E dot full}
\end{eqnarray}
Where the simulations are assumed to span a time in the range of $(0, T)$;
$N_x$, $N_y$, $N_z$ and $N_t$ are number of grid points in the $x$, $y$, $z$ and
time directions respectively; $\mathcal{N}\equiv N_xN_yN_zN_t$,  $\lambda, \mu,
\nu, \rho$ are indices for grid
points in discrete Fourier space: $\mathbf{k}_{\lambda, \mu,
\nu}\equiv\left(2\pi\lambda/L_x, 2\pi\mu/L_y, 2\pi\nu/L_z\right)$,
$w_\rho\equiv2\pi\rho/T$; $\hat{v}$ is the time and
space Discrete Fourier Transform (DFT) of the convective velocity in
the absence of the external perturbation (i.e. the simulated velocity field);
$\hat{\rho}$ is the DFT of the background density along the $z$ direction;
$P_{\lambda,\mu,\nu}\equiv\mathbf{I}- \mathbf{k}_{\lambda,\mu,\nu}
\mathbf{k}_{\lambda,\mu,\nu}/k^2_{\lambda,\mu,\nu}$ is the discrete version of
the $P_k$ operator defined by \citep{Goodman_Oh_97} that maintains
incompressibility.

The $2,2$ subscripts
of $\dot{\mathcal{E}}_{2,2}$ denote that equation \ref{eq: E dot full} contains up to
second order terms in the  two dimensionless parameters characterizing the tide 
and the convection respectively, $\rho(k_z)$ is the Fourier transform
of the density averaged over $x,y,t$. The normalization is such that 
$\rho(0)$ is the average density over all space and time. 

To perform the
calculation we use a discreet Fourier transform to estimate the spectra of the
velocities and density needed in equation \ref{eq: E dot full}, but before that we
again re-sample to a grid that has its collocation points evenly spaced in the
vertical direction. We have to be careful when using discreet Fourier transforms
to estimate spectra. In particular we need to pay special attention to the z and
time directions, since they are not periodic and hence the discreet Fourier
transforms suffer from windowing effects. To confirm that our results are not
significantly affected, we perform the same calculations using no windowing in
those two directions, as well as Welch (parabolic) and Bartlett (triangular) 
windows. Also it is possible
that the impenetrable top and bottom walls might influence our result. So we
also tried a Welch and Bartlett window functions that exclude the top 12.5\% and
the bottom 5\%. The resulting effective viscosity scalings from all these tries
were not significantly different.\\

\begin{figure}
\begin{center}
	\includegraphics[width=0.6\textwidth]{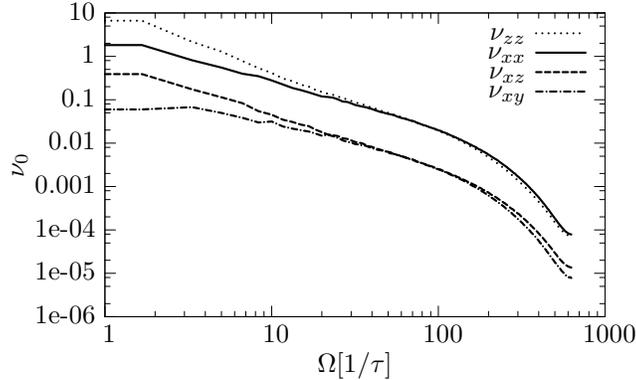}
	\caption{The four independent components of the viscosity tensor as
	estimated by the \citet{Goodman_Oh_97} method. The vertical axis is the
	effective viscosity in units of $<v_z^2>^{1/2}H_p$ ($H_p$ - pressure
	scale height) and the horizontal
	axis is the perturbation frequency is units of inverse convective
	turnover time ($\tau$).}
	\label{fig: nu}
\end{center}
\end{figure}

We then note that the energy transfer rate of equation \ref{eq: E dot full} is due
to the external
forcing. If the flow is steady state this transfer rate must be balanced by the
energy dissipated by the turbulence. We can then extract the different
components of the effective viscosity by setting all terms of $\mathbf{A}$
to zero except for one, and matching to the energy dissipation rate 
that an actual molecular viscosity ($\nu$) would cause:
\begin{equation}
	\dot{\mathcal{E}}_{visc} = \frac{1}{2}
	\left<\rho\nu\right>
	\;Trace\left[\mathbf{A}(\Omega)\cdot\mathbf{A^*}(\Omega)\right]
	\label{eq: E dot mol},
\end{equation}
where averaging is done over the volume of the box and over time. 
Note that this is only the dissipation of the flow caused by the external
shear. In the absence of external shear (i.e. $\mathbf{A}\rightarrow0$) there is
no dissipated energy. The dissipation of the convective flow is not of interest
because it is assumed to be  balanced by the thermal driving of the convection.

Since physically there should be no difference between the two horizontal
directions we expect that $\nu_{xx}\approx\nu_{yy}$, $\nu_{xy}\approx\nu_{yx}$ 
and $\nu_{xz}\approx\nu_{yz}\approx\nu_{zx}\approx\nu_{zy}$. We verified that 
these approximations indeed hold. So we are left with only four independent 
components of the viscosity tensor. Those are presented in figure \ref{fig: nu}.
As we can see, all the components have similar scaling with frequency, and the
diagonal components are a bit over an order of magnitude larger than the 
off--diagonal components. The approximate frequency scaling for the low 
frequency dependence of the viscosity components (for $\Omega<100$) is:
\begin{equation}
	\nu\propto\Omega^{1.2\pm0.1}
	\label{eq: nu scaling}
\end{equation}
\section{Conclusion}
The above result points to the possibility that viscosity in turbulent
convection zones loses efficiency significantly slower than what Kolmogorov 
scaling predicts at least on large timescales. This seems to be due to the fact
that the turbulent eddies with turnover times similar to these large timescales
are not in the inertial subrange and hence, the velocity power spectrum is much 
shallower than the Kolmogorov 5/3 law. This result is not conclusive, since the 
possibility remains that dropping higher order terms in the above expansion is 
not a good approximation. 

In \citet{Penev_Barranco_Sasselov_08b} we introduce horizontal depth dependent
forcing into the flow equations and obtain the dissipation properties of the
turbulent convective flow directly. We were then able to compare the average
rate of work done on the flow
by the external forcing to the expected rate of energy transport and dissipation
by an assumed effective viscosity. We found that with sufficiently long time
average these two quantities have the same depth dependence, thus 
verifying the validity of the assumption that an effective viscosity coefficient
is sufficient to parametrize the average dissipation and momentum transport
properties of the turbulent convective flow. Further, by repeating the above
procedure a number of times we were able to derive the scaling of this effective
viscosity coefficient with period and confirm that it is linear.

In this paper we have presented a code that is well suited for the purpose of
studying turbulent dissipation in convective zones.
First it is a spectral code, which means that the spatial accuracy is
exponential, and hence the code efficient for simulating turbulent flows. 
Further by using the anelastic
approximation we do not need to deal with sound waves and shocks. This allows
us to take much larger time steps (by more than an order of magnitude) than with
a fully compressible code, and hence
we can have runs that cover a much longer physical time than fully compressible
simulations. This is important since the external forcing regimes we are
interested in are tides with orbital periods of several days, which is
not achievable in reasonable time with fully compressible codes. 

The price we pay of course is that the flow we simulate is not a good
approximation to the flow in any star's surface convection zone.
Firstly, we cannot accommodate the region where the driving of the convection
occurs, because this region is characterized by strongly supersonic flows
and shocks, also the mean free path of light in that region is not small and 
hence the radiative effects can no longer be captured by a heat diffusion 
coefficient. Secondly, our code uses the ideal gas equations of state which is a
poor approximation to the upper layers of stars. 

However, the scaling of the effective viscosity with frequency we obtained
(equation \ref{eq: nu scaling}) for 
our box agrees with the scalings obtained with the same perturbative approach
applied to realistic simulations of the upper layers of the convection
zones of the sun and two smaller stars 
\citep{Penev_Sasselov_Robinson_Demarque_07, Penev_Sasselov_Robinson_Demarque_08}.
Which gives us confidence that
our results are applicable to the systems we are interested in studying. 

We would like to acknowledge the contribution of Philip Marcus to the
development of the original code \citep{Barranco_Marcus_06}. 
\appendix
\section{Implementing the Pressure Step}
\label{app: pressure step}
As discussed in section \ref{sec: pressure step} in order to advance the
enthalpy by one time step we need to solve the following differential equation
with boundary conditions:
\begin{eqnarray}
	\left[\nabla^2 + \frac{d\log\bar{\rho}}{dz}\frac{\partial}{\partial
	z}\right] \Pi^{N+1} &=&
	\frac{1}{\Delta t}\left(\mathbf{\nabla}\cdot\mathbf{v}^{N+\frac{2}{3}}+ 
	v_z^{N+\frac{2}{3}}\frac{d\log\bar{\rho}}{dz}\right)\\
	\left.\frac{\partial \Pi^{N+1}}{\partial z}\right|_{\pm L_z/2} &=& 
	\frac{1}{\Delta t}\left.v_z^{N+\frac{2}{3}}\right|_{\pm L_z/2}
	\nonumber
\end{eqnarray}
The solution is obtained in two steps. 
First we impose the boundary conditions by ignoring the differential equation
for the two highest Chebyshev modes and replacing it with the equations for the
boundary conditions. Then we use a Green's step
(see sec. \ref{app: greens step}) to fix the equation for those two highest
modes and instead impose the boundary conditions at the expense of satisfying
the equation at the physical top and bottom boundary.

We need to solve this equation as efficiently as possible. Clearly we can make
this a trivial matrix multiplication operation if we were to store the inverse
of the left hand side operator for each horizontal mode at the start, and at
each time step we multiply every vertical slice of the transformed right hand
side by the corresponding inverse to get the value of $\Pi^{N+1}$. However, this
would require a $N_x\times N_y$ matrices of $N_z^2$ elements to be stored (where
$N_x$, $N_y$ and $N_z$ are the resolutions in the $\hat{x}$, $\hat{y}$ and
$\hat{z}$ directions respectively), which for large resolutions is likely to
exceed the amount of memory available on each node of the cluster where the
code is to run.

To avoid this we need to find the most efficient way which only stores things
common to all horizontal modes. We see that the pressure
equation is almost identical for all horizontal modes, except for the horizontal
component of the $\nabla^2$ operator, which in Fourier space means simply
multiplying by $k_\perp^2\equiv k_x^2 + k_y^2$, where $k_x$ and $k_y$ are the
corresponding $\hat{x}$ and $\hat{y}$ wavenumbers for each mode. So what we can
reasonably do is pre-compute the common part of the left hand side operator and
then for each horizontal mode add $k_\perp^2$ along the diagonal. We then
overwrite the last row of the resulting matrix ($\mathbf{M}_i,j$) with
\begin{eqnarray}
	M_{N_z,p}&=&p^2\\
	M_{N_z-1,p}&=&(-1)^p p^2 
\end{eqnarray}
in order to impose the boundary conditions. We then decompose $\mathbf{M}$ into 
an upper and lower triangular parts and solve by backward substitution. 
The right hand side vector
also needs to have its highest two entries overwritten with the boundary
conditions at the top and bottom of the box respectively. 

\subsection{Green's Step}
\label{app: greens step}
In this step we fix the anelastic constraint even for the two highest Chebyshev
modes and instead break the two highest modes of the momentum equation to satisfy 
the boundary conditions which we break in the process. To make expressions
shorter define the following operators:
\begin{eqnarray}
	D &\equiv& \frac{d}{dz} \label{eq: D def}\\
	D_A &\equiv& \frac{d}{dz}+\frac{d\ln \bar{\rho}}{dz} 
							\label{eq: D_A def}\\
	\mathbf{\nabla} &\equiv& ik_x \mathbf{\hat{x}}+ik_y \mathbf{\hat{y}}
		+ D\mathbf{\hat{z}} \label{eq: nabla def}\\
	\mathbf{\nabla}_A &\equiv& ik_x \mathbf{\hat{x}}+ik_y \mathbf{\hat{y}}
		+ D_A\mathbf{\hat{z}} \label{eq: nabla_A def}\\
	\Delta_A &\equiv& \mathbf{\nabla}_A\cdot\mathbf{\nabla}
							\label{eq: Delta_A def}
\end{eqnarray}
We would like to modify the pressure step in a way that will include two new
degrees of freedom which we can then use to fix the anelastic constraint for the
two highest Chebyshev modes. The particular modification useful in this case is:
\begin{equation}
	\mathbf{v}^{N+1} = 
	\mathbf{v}^{N+\frac{2}{3}} - \Delta t\mathbf{\nabla}\Pi^{N+1} +
	\mathbf{\hat{z}} \left(\tau_1^{N+1} T_{M-1} + \tau_2^{N+1} T_M\right)
\end{equation}
Where $T_N$ denotes a Chebyshev polynomial of order $N$, $\tau_1$ and $\tau_2$
are arbitrary coefficients to be chosen later and $M$ is the order of the 
highest Chebyshev coefficient we are simulating.

Requiring the anelastic constraint and velocity boundary conditions gives:
\begin{eqnarray}
	 \Delta_A \Pi^{N+1} &=&
	\frac{1}{\Delta t} \mathbf{\nabla}_A\cdot\mathbf{v}^{N+\frac{2}{3}} +
	\tau_1^{N+1} D_A T_{M-1} + \tau_2^{N+1} D_A T_M\\
	D\left.\Pi^{N+1}\right|_{\pm L_z/2} &=& 
	\frac{1}{\Delta t}\left.v_z^{N+\frac{2}{3}} +
	\tau_1^{N+1} T_{M-1} + \tau_2^{N+1} T_M \right|_{\pm L_z/2}
\end{eqnarray}
To proceed we break up $\Pi^{N+1}$ into 3 pieces:
\begin{equation}
	\Pi^{N+1}=\Pi_0^{N+1} + \frac{1}{\Delta t} \left(
	\tau_1^{N+1}\Gamma_1^{N+1} + \tau_2^{N+1}
	\Gamma_2^{N+1}\right)
\end{equation}
This allows us to split the above equation into 3 separate equations with
corresponding boundary conditions, the first of which is the already calculated 
pressure step:
\begin{eqnarray}
	\Delta_A \Pi_0^{N+1} &=&
	\frac{1}{\Delta t} \mathbf{\nabla}_A\cdot\mathbf{v}^{N+\frac{2}{3}} \\
	D\left.\Pi^{N+1}\right|_{\pm L_z/2} &=& 
	\frac{1}{\Delta t}\left.v_z^{N+\frac{2}{3}}\right|_{\pm L_z/2} \\
	\Delta_A \Gamma_1^{N+1} &=& D_A T_{M-1}\\
	\left.D\Gamma_1^{N+1}\right|_{\pm L_z/2} &=& \left.T_{M-1}\right|_{\pm
	L_z/2}\\
	\Delta_A \Gamma_2^{N+1} &=& D_A T_M\\
	\left.D\Gamma_2^{N+1}\right|_{\pm L_z/2} &=& \left.T_M\right|_{\pm
	L_z/2}
\end{eqnarray}
We solve the two new equations the same way we solve the first. This makes as
before the anelastic constraint satisfied for all but the two highest Chebyshev
modes, but this time we have 2 arbitrary constants $\tau_1$ and $\tau_2$ which
we can then set to values that will make the anelastic constraint hold for 
those two modes as well. 

\section{Implementing the Heat Diffusion Step}
\label{app: heat diffusion}
The same considerations as the pressure step apply to this step. We again decide
against making a table of pre-inverted matrices for each horizontal mode, and 
instead we only pre-compute the part of the matrix that is common for all
horizontal modes. To make the description of the numerical procedure followed 
we define the following matrices:\\
\begin{tabular}{c@{:}l}
	$CP$&transforms a vector from Chebyshev to physical space\\
	$PC$&transforms a vector from physical to Chebyshev space\\
	$D$&differentiates a vector in Chebyshev space\\
\end{tabular}\\
From those we construct a derivative operator and the common part of the 
left hand side operator, both in physical space:
\begin{eqnarray}
	D_P &\equiv& CP\cdot D\cdot PC \\
	M_P &\equiv& D_P\cdot D_P + F\cdot D_P + G
\end{eqnarray}

What we pre-compute and store is a matrix $\mathbf{M}$, which we obtain by
overwriting the first and last row of $M_P$ with:
\begin{eqnarray}
	{M_P}_{1,1}=1\quad&& {M_P}_{1,j}=0,\ j=2..N_z\\
	{M_P}_{N_z, j}=0,\ j=1..N_z-1\quad&& {M_P}_{N_z, N_z}=1
\end{eqnarray}
to allow for imposing the boundary conditions, then sandwiching the resulting
matrix between $PC$ and $CP$:
\begin{equation}
	\mathbf{M}\equiv PC \cdot M_P \cdot CP
\end{equation}

Then in order to solve equations \ref{eq: heat diff} and \ref{eq: heat diff bc}
for each horizontal mode we construct a new matrix $\mathbf{M}'(k_\perp)$ 
from $\mathbf{M}$ as follows:
\begin{equation}
	\mathbf{M}'_{i,j} (k_\perp) \equiv \mathbf{M}_{i,j} - k_\perp^2 \delta_{i,j} 
	+ k_\perp^2 \left(PC_{i,0} CP_{0,j} + PC_{i,N_z-1} CP_{N_z-1,j}\right)
\end{equation}
The first new term adds the horizontal part of the $\nabla^2$ operator 
for the given mode. However, this breaks the requirements for the boundary 
conditions, so we repair them with the other two terms. This way the boundary
conditions are directly imposed by breaking equation \ref{eq: heat diff} at the
physical top and bottom of the box, instead of ignoring it for the two highest
Chebyshev modes. That means we do not need to perform extra greens steps like 
we did for the pressure equation. Then we obtain the solution to equation 
\ref{eq: heat diff} by decomposing $\mathbf{M}'(k_\perp)$ into an upper and
lower triangular matrices and using back-substitution.

\end{document}